\documentclass[12pt]{article}
\usepackage{a4}
\usepackage{amsmath}
\usepackage{amsfonts}
\usepackage{amsmath}
\usepackage{latexsym}
\usepackage{amsfonts}

\renewcommand{\theequation}{\arabic{section}.\arabic{equation}}
\begin{document}
	
	\noindent    \hfill  April    2021 \\
	
	\noindent \vskip 2cm
	
	\begin{center}
		
		{\Large\bf On special quartic interaction of higher spin gauge fields with scalars and gauge symmetry  commutator in the linear approximation}
		\bigskip\bigskip\bigskip
	\end{center}
	
	\vspace*{1 cm}
	\begin{center}
		{\large Melik Karapetyan, Ruben Manvelyan and Gabriel Poghosyan}
		
		\medskip
		
		{\small\it A.Alikhanyan National Laboratory (Yerevan
			Physics Institute)\\ Alikhanian Br. Str. 2, 0036 Yerevan, Armenia}\\
		\medskip
		{\small\tt meliq.karapetyan@gmail.com; manvel@yerphi.am; gabrielpoghos@gmail.com}
	\end{center}
	\vspace{2cm}

	\bigskip
	\begin{center}
		{\sc Abstract}
	\end{center}
	\quad
	Local quartic interaction of higher-spin gauge field with a scalar field is considered. In this special case, the nontrivial task of construction of interacting Lagrangian for the higher spin field in physical gauge was solved using the full power of Noether's procedure. As a result, the linear on-field gauge transformation is obtained and the corresponding commutator of transformation is analyzed. To understand the closure of this algebra the right-hand side of this commutator is classified in respect to gauge transformations coming from cubic interactions with different higher spin symmetric tensor fields and with mixed symmetry tensor fields transformations.

\section*{Introduction}
	
	\quad The problem of construction of complete Higher Spin (HS) Interaction Lagrangian is one of the tasks with smoldering interest for last forty years \cite{VasilievEqn}-\cite{Sleight:2017krf}. From time to time this complicated and puzzling problem came to a relative center of interest when some even small progress was observed in interaction construction or when the HS theories start to play some role in the development of other important fields like AdS/CFT.
	Further, we can say that this task is interesting in itself due to the complexity and the necessity to develop and use nontrivial computing techniques even for small advances and achievements.
	
	Even though during the last ten-twelve years we observed significant progress in the understanding of construction and structure of cubic interaction in different approaches, dimensions and backgrounds \cite{Bengtsson:1983pd}-\cite{Khabarov:2020bgr}, our knowledge about higher-order vertices is far from complete (see, however, \cite{Ruehl:2011tk}-\cite{Roiban:2017iqg})  ]) and seems to be bounded by the idea that the quartic one should be nonlocal in general \cite{Bengtsson:2016hss}-\cite{Joung:2019wbl}.
	We should mention also that all these activities supplemented with the parallel development of Vasiliev’s nonlinear theory of interacting HS field equations of motion in AdS background, where the questions of possible non-locality beyond the cubic level also arose and discussed (for last development see \cite{Didenko:2020bxd}, \cite{Gelfond:2019tac} and ref. there).
	Nevertheless, it seems that in some special cases it is possible to construct some local interactions between fields with different spins at least as a part of a more complicated covering theory (maybe non-local)  including other gauge fields and symmetries. It is appropriate here to mention the existence of a complete, local and one-loop finite chiral theory of higher spins in four-dimensional flat space, constructed and developed in \cite{Ponomarev:2016lrm}-\cite{Skvortsov:2020gpn}.
	
	In this paper, we construct some special local quartic interaction of two scalars and two spin four fields using standard Noether's procedure. The interesting points of this special construction are the following:
	\begin{itemize}
		\item First we see that to close  Noether's procedure we should add additional cubic interaction of scalar with other spin gauge fields and corresponding HS gauge symmetries.
		\item Second important point that we constructing quartic vertex we derive fixed linear in gauge field gauge transformation of our HS field $\delta_{1}^{(\epsilon)}$ and then be able to investigate the closure of commutators of two such a transformation
		\begin{equation}\nonumber
		[\delta_{1}^{(\eta)}\delta_{1}^{(\epsilon)}]\sim \delta_{1}^{([\eta,\epsilon])} + \textnormal{additional terms}
		\end{equation}
		and understand whether it leads to nonlocality or not.
	\end{itemize}
	We organize our paper in the following way: in the next section, we show spin 2 exercise for construction similar quartic interaction in spin-two case. Then the essential Noether's construction is shown in section two with the final derivation of
	interaction Lagrangian and first order on HS gauge field gauge transformation. The last section is devoted to the investigation of the commutator of two $\delta_{1}$ transformations and classification of terms different from the same $\delta_{1}$ with the composed parameter on the right-hand side.
	The calculations which we have performed here are based on the technique and notation developed in the past in \cite{MR}-\cite{deWit:1979sib}.
	
\section{Illustration: Spin two case}
	\setcounter{equation}{0}
	\quad For better understanding how we should construct special quartic interaction in higher spin case we start first from the following lagrangian for the interaction of scalar with spin 2 gauge field in the flat background:
	\begin{equation}\label{1.1}
	S^{\Phi\Phi  h^{(2)}}=S_{0}(\Phi)+ S_{1}(\Phi,h^{(2)}) ,
	\end{equation}
	where
	\begin{eqnarray}
	S_{0}(\Phi)&=&\frac{1}{2}\int d^{d}x\partial_{\mu}\Phi\partial^{\mu}\Phi, \label{1.2}\\
	S_{1}(\Phi,h^{(2)})&=&\frac{1}{2}\int d^{d}x h^{(2)\mu\nu}\Big[-\partial_{\mu}\Phi\partial_{\nu}\Phi
	+\frac{\eta_{\mu\nu}}{2}\partial_{\lambda}\Phi\partial^{\lambda}\Phi\Big] .\quad\label{1.3}
	\end{eqnarray}
	This is a well known minimal coupling of scalar with gravity, linearized in flat background and the bracket in (\ref{1.3}) is the usual energy-momentum tensor for massless scalar field. The spliting  of  (\ref{1.1}) into the quadratic and cubic
	parts allows us to formulate the Noether's equations
	\begin{equation}\label{1.4}
	\delta_{1}S_{0}(\Phi)+\delta_{0}S_{1}(\Phi,h^{(2)})=0 ,
	\end{equation}
	where:
	\begin{eqnarray}
	\delta_{0} h^{(2)}_{\mu\nu} &=& \partial_{(\mu}\varepsilon^{(1)}_{\nu)}=\partial_{\mu}\varepsilon^{(1)}_{\nu}+\partial_{\nu}\varepsilon^{(1)}_{\mu}, \label{1.5} \\
	\delta_{0} \Phi &=& 0 , \label{1.6}\\
	\delta_{1}\Phi &=&\varepsilon^{{(1)}\lambda}\partial_{\lambda}\Phi .\label{1.7}
	\end{eqnarray}
	
	The crucial point here that we can discover interacting part (\ref{1.3}) solving functional equation (\ref{1.4}) variating known free part (\ref{1.2})  in respect to admitting first order diffeomorphism of scalar (\ref{1.7}) and using a zero-order variation of gauge field as an integration rule. Note also that the scalar field has no zero-order variation being matter field here.
	
	Then we can formulate task for construction of the next order interaction of the two scalars with two spin two fields in the similar  way:
	\begin{equation}\label{1.8}
	\delta_{2}S_{0}(\Phi) + \delta_{1}S_{1}(\Phi, h^{(2)})+\delta_{0}S_{2}(\Phi, h^{(2)})=0
	\end{equation}
	admitting that: $\delta_{2}\Phi=0$ we see that in this case we need to solve again two terms functional equation
	\begin{equation}\label{1.9}
	\delta_{1}S_{1}(\Phi, h^{(2)})+\delta_{0}S_{2}(\Phi, h^{(2)})=0 ,
	\end{equation}
	using the same transformations (\ref{1.5})-(\ref{1.7}) and introducing an assumption about the form of first order transformation of spin 2 gauge field:
	\begin{eqnarray}
	\delta_{1} h^{(1)}_{\mu\nu} &=& \varepsilon^{(1)\lambda}\partial_{\lambda}h^{(2)}_{\mu\nu}+\bar{\delta_{1}} h^{(2)}_{\mu\nu} ,\label{1.10}
	\end{eqnarray}
	where $\bar{\delta_{1}} h_{\mu\nu}$ we should find from equation (\ref{1.9}) together with $S_{2}(\Phi, h^{(2)})$.
	To show technology of solution we present variation of $S_{1}(\Phi, h^{(2)})$ in the following form (after some algebra and partial integrations)
	\begin{eqnarray}
	\delta_{1}S_{1}(\Phi, h^{(2)})&=&\int d^{d}x \Big\{ -\frac{1}{2}\partial^{\mu}\Phi\partial^{\nu}\Phi
	[\bar{\delta_{1}} h^{(2)}_{\mu\nu}-\partial_{(\mu}\varepsilon^{(1)\lambda}h^{(2)}_{\nu)\lambda}+2 h^{(2)\lambda}_{\mu}\partial_{(\nu}\varepsilon^{(1)}_{\lambda)}]\quad\quad \nonumber\\
	&+&\frac{1}{2}\partial^{\mu}\Phi\partial^{\nu}\Phi[\partial_{\lambda}\varepsilon^{(1)\lambda}h^{(2)}_{\mu\nu}
	+\frac{1}{2}\partial_{(\mu}\varepsilon^{(1)}_{\nu)}h^{(2)\alpha}_{\alpha}]\nonumber\\
	&+&\frac{1}{4}\partial^{\lambda}\Phi\partial_{\lambda}\Phi[\bar{\delta_{1}}h^{(2)\alpha}_{\alpha} - \partial_{\beta}\varepsilon^{(1)\beta}h^{(2)\alpha}_{\alpha}]\Big\} .\label{1.11}
	\end{eqnarray}
	From first line of (\ref{1.11}) we can derive that if we define
	\begin{equation}\label{1.12}
	\bar{\delta_{1}} h^{(2)}_{\mu\nu}=\partial_{(\mu}\varepsilon^{(1)\lambda}h^{(2)}_{\nu)\lambda} ,
	\end{equation}
	then last term of first line can be integrated to $\delta_{0}[-\frac{1}{2}\partial^{\mu}\Phi\partial^{\nu}\Phi h^{\lambda}_{\mu}h_{\nu\lambda}]$
	
	Then taking into account that
	\begin{eqnarray}
	\bar{\delta}_{0}h^{(2)\alpha}_{\alpha} &=& 2\partial_{\lambda}\varepsilon^{(1)\lambda},\label{1.13} \\
	\bar{\delta}_{1}h^{(2)\alpha}_{\alpha}&=& 2 \partial^{\lambda}\varepsilon^{(1)\alpha}h^{(2)}_{\lambda\alpha}=\frac{1}{2}\delta_{0}[h^{(2)\lambda\alpha}h^{(2)}_{\lambda\alpha}] ,\label{1.14}
	\end{eqnarray}
	we see that we can immediately integrate second and third line of (\ref{1.11}) and arrive to the following quartic action:
	\begin{eqnarray}
	S_{2}(\Phi, h^{(2)})&=&\int d^{d}x \Big\{ \frac{1}{2}\partial^{\mu}\Phi\partial^{\nu}\Phi
	h^{(2)\lambda}_{\mu}h^{(2)}_{\nu\lambda}-\frac{1}{4}\partial^{\mu}\Phi\partial^{\nu}\Phi h^{(2)}_{\mu\nu}h^{(2)\alpha}_{\alpha}\nonumber\\
	&-&\frac{1}{8}\partial^{\lambda}\Phi\partial_{\lambda}\Phi h^{(2)\alpha\beta}h^{(2)}_{\alpha\beta} + \frac{1}{16}\partial^{\lambda}\Phi\partial_{\lambda}\Phi h^{(2)\alpha}_{\alpha}h^{(2)\beta}_{\beta}\Big\} ,\label{1.15}
	\end{eqnarray}
	with expected Lie derivative as a solution for first order variation of spin two fluctuation:
	\begin{equation}\label{1.16}
	\delta^{(\varepsilon)}_{1}h^{(2)}_{\mu\nu}=\varepsilon^{(1)\lambda}\partial_{\lambda}h^{(2)}_{\mu\nu}+\partial_{\mu}
	\varepsilon^{(1)\lambda}h^{(2)}_{\nu\lambda}+\partial_{\nu}\varepsilon^{(1)\lambda}h^{(2)}_{\mu\lambda}
	=\mathfrak{L}_{\varepsilon^{(1)\lambda}}h^{(2)}_{\mu\nu} ,
	\end{equation}
	with standard algebra
	\begin{equation}\label{1.17}
	[\delta^{(\eta)}_{1},\delta^{(\varepsilon)}_{1}]h^{(2)}_{\mu\nu}=
	\delta^{([\varepsilon,\eta])}_{1}h^{(2)}_{\mu\nu} ,
	\end{equation}
	where composite parameter is the usual Lie commutator of vectors:
	\begin{equation}\label{1.18}
	[\eta,\varepsilon]=[\eta^{(1)},\varepsilon^{(1)}]^{\lambda}
	=\eta^{(1)\nu}\partial_{\nu}\varepsilon^{(1)\lambda}
	-\varepsilon^{(1)\nu}\partial_{\nu}\eta^{(1)\lambda} .
	\end{equation}
	Note that one can work with restricted external field also. As an example we can choose at once traceless $h^{(2)\mu\nu}$ supplemented with the corresponding constraint on gauge parameter $\partial_{\mu}\varepsilon^{(1)\mu}=0$. This means that in expression (\ref{1.11}) survive only first line and first term of the last line. Then we see from (\ref{1.12}) and (\ref{1.14}) that $\bar{\delta_{1}} h^{(2)}_{\mu\nu}$ is not traceless but can be integrated. So we arrive to the first and third terms of interaction (\ref{1.15}) describing interaction of $h^{(2)\mu\lambda}h^{(2)\nu}_{\lambda}$ with traceless part of current $J^{(2)}_{\mu\nu}=\partial_{\mu}\Phi\partial_{\nu}\Phi$.
	
	\section{Spin four case}
	
\subsection*{Setup}
	\setcounter{equation}{0}
	Our main task is to construct similar quartic interaction for spin 4 using prescriptions developed in the previous simple spin 2 case. In our previous articles \cite{MR}, \cite{MM} we prove that in both $AdS$ and flat backgrounds after corresponding field redefinition interaction of even spin $s$  gauge field with spin $s$ current constructed from scalar and derivatives could be written in the form supplemented by the whole tower of invariant actions for couplings of the same scalar with all gauge fields of smaller even spin. So the starting lagrangian for our task we take from \cite{MM} rewriting all terms in the flat background:
	\begin{eqnarray}
	S^{\Phi\Phi h^{(4)}}(\Phi,h^{(2)},h^{(4)})&=&S_{0}(\Phi)+S_{1}(\Phi,h^{(2)})+S_{1}(\Phi,h^{(4)})\label{2.1},
	\end{eqnarray}
	where $S_{0}(\Phi)$ , $S_{1}(\Phi,h^{(2)})$ are defined in (\ref{1.1})-(\ref{1.3}) and
	\begin{eqnarray}
	&&S_{1}(\Phi,h^{(4)})=\frac{1}{4}\int d^{d}x h^{(4)\mu\nu\alpha\beta}[\partial_{\mu}
	\partial_{\nu}\Phi\partial_{\alpha}\partial_{\beta}\Phi-\eta_{\mu\nu}\partial_{\alpha}
	\partial^{\gamma}\Phi\partial_{\beta}\partial_{\gamma}\Phi].\label{2.2}
	\end{eqnarray}
	
	From now on to avoid cumbersome notation and overlapping with symmetrization brackets we reserve notation $h$ and $\varepsilon$ for gauge field for spin four $h^{(4)}$ and corresponding gauge parameter $\varepsilon^{(3)}$ except the cases when we do not explicitly write out indices. In the case of other spin (rank) fields and parameters, we use these letters with an exact indication of rank.
	
	The action (\ref{2.1}) is invariant with respect to the gauge transformations of the spin four field with  an additional  spin two field gauge transformation inspired by the second divergence of the spin four gauge parameter\footnote{Note that the spin two part of our action continues to be invariant in respect of usual linearized  reparametrization}
	\begin{eqnarray}
	\delta_{1}\Phi(x)&=&\varepsilon^{\mu\nu\lambda}(x)\partial_{\mu}\partial_{\nu}\partial_{\lambda}\Phi(x)
	,\label{2.3}\\
	\delta_0 h^{\mu \nu \lambda \rho} &=& \partial^{( \mu} \varepsilon^{\nu \lambda \rho )} = \partial^{ \mu} \varepsilon^{\nu \lambda \rho} + \partial^{\nu } \varepsilon^{\mu \lambda \rho} + \partial^{\lambda } \varepsilon^{\mu \nu \rho} + \partial^{\rho } \varepsilon^{\mu \nu \lambda },\label{2.4}\\
	\delta_{0}h_{(2)}^{\mu\nu}&=&\partial^{(\mu}\epsilon^{\nu)},\label{2.5}\\
	\epsilon^{\nu}&=&\partial_{\alpha}\partial_{\beta}\varepsilon^{\nu\alpha\beta}.\label{2.6}
	\end{eqnarray}
	For further simplification in calculation of quartic terms in spin four case we will use physical traceless and transverse  gauge for our external spin four field:
	\begin{eqnarray}
	\partial_{\mu}h^{\mu\nu\lambda\rho} &=& 0 , \label{2.7}\\
	h_{\mu}^{\mu\lambda\rho}&=& 0 ,\label{2.8}
	\end{eqnarray}
	which leads to the corresponding restrictions on already traceless  spin four gauge parameter:
	\begin{eqnarray}
	\partial_{\alpha}\varepsilon^{\alpha\beta\gamma} &=& 0 , \label{2.9}\\
	\partial_{\mu}\partial^{\mu}\varepsilon^{\alpha\beta\gamma}&=&\Box \varepsilon^{\alpha\beta\gamma}= 0 .\label{2.10}
	\end{eqnarray}
	Note that because in our gauge the gauge parameter is transverse, we should get decoupling of spin two mode from spin four due to degeneration of the additional gauge transformation (\ref{2.5}). Another convention is that
	from now on we will admit integration everywhere where it is necessary. So we work with a
	Lagrangian as with action and therefore we neglect all $d$ dimensional space-time total
	derivatives when making a partial integration.
	
\subsection*{Variation of cubic term}
	So we arrive to the following simplified task: Starting from a single cubic term due to (\ref{2.7})-(\ref{2.10})\footnote{As usual in our articles we widely use Lagrangian instead of Action performing Noether procedure admitting possibility for partial integration}
	\begin{equation}\label{3.1}
	L_{1}\sim h^{\mu \nu \lambda \rho} \partial_\mu \partial_\nu \Phi \partial_\lambda \partial_\rho \Phi ,
	\end{equation}
	and using known variation:
	\begin{eqnarray}
	\delta_1 \Phi &=& \varepsilon^{\alpha \beta \gamma} \partial_\alpha \partial_\beta \partial_\gamma \Phi ,\label{3.2}\\
	\delta_0 h^{\mu \nu \lambda \rho} &=& \partial^{( \mu} \varepsilon^{\nu \lambda \rho )} ,\label{3.3}
	\end{eqnarray}
	we try to solve functional equation:
	\begin{equation}\label{3.4}
	\delta _1 L_1(\Phi, h^{(4)}) + \delta_0 L_2(\Phi, h^{(4)}) =0 ,
	\end{equation}
	and construct unknown quartic interaction and first order gauge variation of spin four field $\delta_{1}h^{\mu\nu\lambda\rho}$. Doing that and taking into account that according to (\ref{2.7}) and (\ref{2.9}) $\alpha,\beta,\gamma$ derivatives commute with $\varepsilon$ and $\mu,\nu,\lambda,\rho$ derivatives commute with $h$ and after long manipulations and multiple partial integrations we arrive to the following important variation:
	\begin{eqnarray}
	\delta_1 (h^{\mu \nu \lambda \rho} \partial_\mu \partial_\nu \Phi \partial_\lambda \partial_\rho \Phi)&=&\frac{1}{3}\delta_1 (h^{\mu \nu \lambda \rho}J^{(4)}_{\mu\nu\lambda\rho})=\delta_1 h^{\mu \nu \lambda \rho} \partial_\mu \partial_\nu \Phi \partial_\lambda \partial_\rho \Phi\nonumber\\
	&+&\frac{1}{50}\left[\varepsilon^{\mu (\alpha\beta} \partial_\mu h^{\gamma \nu \lambda \rho)}-\partial_\mu \varepsilon^{(\alpha \beta \gamma} h^{\nu \lambda \rho )\mu}\right] J^{(6)}_{\nu \lambda \rho \alpha \beta \gamma}\nonumber\\
	&+& \frac{1}{5} \left[\partial_\alpha   \varepsilon^{\mu\nu (\beta} \partial_\mu  \partial_\nu h^{\gamma \lambda \rho ) \alpha } -\partial_\mu \partial_\nu \varepsilon^{\alpha ( \beta \gamma} \partial_\alpha h^{ \lambda \rho ) \mu \nu}\right]  J^{(4)}_{\lambda \rho \beta \gamma}\nonumber\\
	&+&\frac{2}{15}  \left[\partial_\alpha \partial_\beta \partial_\gamma \varepsilon^{(\mu\nu\lambda} h^{\rho ) \alpha\beta\gamma }-\varepsilon^{\alpha \beta \gamma} \partial_\alpha \partial_\beta \partial_\gamma  h^{\mu \nu \lambda \rho}\right] J^{(4)}_{\mu \nu \lambda \rho}\nonumber\\
	&+&\frac{1}{5}[\partial_\mu \partial_\nu \varepsilon^{\alpha \beta \gamma} \partial_\alpha \partial_\beta \partial_\gamma h^{\mu \nu \lambda \rho}
	-\partial_\mu \partial_\nu \partial_\gamma \varepsilon^{\alpha \beta ( \lambda} \partial_\alpha \partial_\beta h^{\rho )\mu \nu \gamma }] J^{(2)}_{\lambda\rho}\nonumber\\
	&+&\frac{1}{5}\partial_\mu \partial_\nu \partial_\lambda \partial_\rho  \varepsilon^{\alpha \beta \gamma} \partial_\alpha  h^{\mu \nu \lambda \rho} J^{(2)}_{\beta \gamma} .\quad\quad\quad \label{3.5}
	\end{eqnarray}
	Here  $J^{(6)}, J^{(4)}, J^{(2)}$ are symmetrized currents:
	\begin{eqnarray}
	J^{(6)}_{\nu \lambda \rho \alpha \beta\gamma} &=& \partial_{(\nu}\partial_{\lambda}\partial_{\rho}\Phi\partial_{\alpha}\partial_{\beta}\partial_{\gamma)}\Phi= \partial_{\nu}\partial_{(\lambda}\partial_{\rho}\Phi\partial_{\alpha}\partial_{\beta}\partial_{\gamma)}\Phi,\quad  \label{3.6}\\
	J^{(4)}_{\mu \nu \lambda \rho} &=&\partial_{(\mu}\partial_{\nu}\Phi\partial_{\lambda}\partial_{\rho)}\Phi= \partial_{\mu}\partial_{(\nu}\Phi\partial_{\lambda}\partial_{\rho)}\Phi ,\label{3.7}\\
	J^{(2)}_{\mu \nu} &=& \partial_{\mu}\Phi\partial_{\nu}\Phi .  \label{3.8}
	\end{eqnarray}
	From (\ref{3.5}) we see several differences from spin  two case:
	\begin{itemize}
		\item From second line of (\ref{3.5}) follows that \emph{we cannot integrate Noether's equation without introduction of the cubic interaction with a gauge field of spin 6 coupled to the spin 6 current:}
		\begin{equation}\label{3.9}
		h_{(6)}^{\nu \lambda \rho \alpha \beta\gamma}J^{(6)}_{\nu \lambda \rho \alpha \beta\gamma} .
		\end{equation}
		\item From third  and fourth lines we see that $J^{(4)}$ terms  arose with different weight $\frac{1}{5}$ and $\frac{2}{15}$. But we will see below that they should come with same weight  to complete integration for interaction terms.
		\item In last two lines we have three unwanted $J^{(2)}$ terms. We should discover way to get rid of them.
	\end{itemize}

	To remove these three obstructions we note that there are several connections between our parts in (\ref{3.5}) leading to
	a redefinition of the initial cubic interactions.
	In another words we can modify our initial interaction with higher spin currents adding gradients of lower spin currents with some coefficients:
	\begin{eqnarray}
	J^{(6)}_{\alpha\beta\mu\nu\lambda\rho} &=>& J^{(6)}_{\alpha\beta\mu\nu\lambda\rho}+A \partial_{(\alpha}\partial_{\beta}J^{(4)}_{\mu\nu\lambda\rho)}+ B \partial_{(\alpha}\partial_{\beta}\partial_{\mu}\partial_{\nu} J^{(2)}_{\lambda\rho)} ,\label{4.1}\\
	J^{(4)}_{\mu\nu\lambda\rho} &=>& J^{(4)}_{\mu\nu\lambda\rho}+ C \partial_{(\mu}\partial_{\nu} J^{(2)}_{\lambda\rho)} .\label{4.2}
	\end{eqnarray}
	\emph{And it works!} Hiding all details of derivations in Appendix A we present final variation  we obtained by tuning procedure (\ref{4.1}) instead of (\ref{3.5})
	\begin{eqnarray}
	\delta_1 (h^{\mu \nu \lambda \rho} \partial_\mu \partial_\nu \Phi \partial_\lambda \partial_\rho \Phi)&=&\frac{1}{3}\delta_1 (h^{\mu \nu \lambda \rho}J^{(4)}_{\mu\nu\lambda\rho})=\delta_1 h^{\mu \nu \lambda \rho} \partial_\mu \partial_\nu \Phi \partial_\lambda \partial_\rho \Phi\nonumber\\
	&+&\frac{1}{50}\left[\varepsilon^{\mu (\alpha\beta} \partial_\mu h^{\gamma \nu \lambda \rho)}-\partial_\mu \varepsilon^{(\alpha \beta \gamma} h^{\nu \lambda \rho )\mu}\right] \tilde{J}^{(6)}_{\nu \lambda \rho \alpha \beta \gamma}\nonumber\\
	&+& \frac{1}{6} \left[\partial_\alpha   \varepsilon^{\mu\nu (\beta} \partial_\mu  \partial_\nu h^{\gamma \lambda \rho ) \alpha } -\partial_\mu \partial_\nu \varepsilon^{\alpha ( \beta \gamma} \partial_\alpha h^{ \lambda \rho ) \mu \nu}\right]  J^{(4)}_{\lambda \rho \beta \gamma}\nonumber\\
	&+&\frac{1}{6}  \left[\partial_\alpha \partial_\beta \partial_\gamma \varepsilon^{(\mu\nu\lambda} h^{\rho ) \alpha\beta\gamma }-\varepsilon^{\alpha \beta \gamma} \partial_\alpha \partial_\beta \partial_\gamma  h^{\mu \nu \lambda \rho}\right] J^{(4)}_{\mu \nu \lambda , \rho}\nonumber\\\label{4.5last}
	\end{eqnarray}
	where modified $\tilde{J}^{(6)}$ is
	\begin{eqnarray}\label{4.6last}
	\tilde{J}^{(6)}_{\nu \lambda \rho \alpha \beta \gamma}=J^{(6)}_{\nu \lambda \rho \alpha \beta \gamma}+\frac{1}{9} \partial_{(\alpha} \partial_\beta J^{(4)}_{ \gamma \nu \lambda \rho )}+\frac{1}{3} \partial_{(\nu} \partial_\lambda \partial_\rho \partial_\alpha  J^{(2)}_{\beta \gamma)}.
	\end{eqnarray}
	Supplemented by traceless Stueckelberg like transformation (\ref{4.9}) of the spin two gauge field from linear coupling with $J^{(2)}$ current:
	\begin{equation}\label{4.9last}
	\delta_{1}h_{(2)}^{\beta \gamma} \sim \partial_\mu \partial_\nu \partial_\lambda \partial_\rho  \varepsilon^{\alpha \beta \gamma} \partial_\alpha  h^{\mu \nu \lambda \rho}.
	\end{equation}
	
	\subsection*{Integration and interaction}

	Now we can start to integrate the last three lines of expression (\ref{4.5last}). Doing that in the corresponding subsection of Appendix A we finally obtain quartic interactions:
	\begin{gather}
	S_{2}(\Phi, h^{(4)})=\int d^{d}x \Big\{\frac{1}{10} h^{\alpha \beta \gamma}_\mu h^{\nu \lambda \rho \mu}\tilde{J}^{(6)}_{\nu \lambda \rho \alpha \beta \gamma}\nonumber\\ - \frac{2}{3} h^{\alpha \beta \gamma}_{\mu} \partial_\alpha  \partial_\beta h^{\mu\nu\lambda\rho}J^{(4)}_{\nu \lambda \rho \gamma}
	+\frac{1}{2} \partial_{\nu} h^{\alpha  \beta \gamma}_{\mu} \partial_\alpha h^{ \mu\nu\lambda\rho} J^{(4)}_{\lambda \rho \beta \gamma}
	-\frac{1}{4} \partial^{\alpha}  h^{\beta \gamma }_{\mu \nu} \partial_\alpha h^{\mu\nu\lambda\rho} J^{(4)}_{\lambda \rho \beta \gamma}\nonumber \\
	- \partial^{\beta}  h^{\alpha \gamma }_{\mu \nu} \partial_\alpha h^{\mu\nu\lambda\rho}J^{(4)}_{\lambda \rho \beta \gamma}
	+\frac{1}{3} \partial^\beta  h^{\gamma}_{\mu \nu \lambda} \partial^{\alpha}  h^{\mu\nu\lambda\rho}J^{(4)}_{\rho \alpha \beta \gamma}-\frac{1}{4}h^{\beta \gamma }_{\mu \nu} h^{\lambda \rho \mu \nu}\Box  J^{(4)}_{\lambda \rho \beta \gamma}\Big\} ,\label{5.7last}
	\end{gather}
	and linear on spin four  gauge field transformations fixed by Noether's procedure:
	\begin{gather}
	\delta_1 h_{(6)}^{\mu \nu \lambda \alpha \beta \gamma}=\varepsilon^{\rho (\alpha \beta}\partial_{\rho}h^{\gamma \mu \nu \lambda )}+\partial^{(\alpha}\varepsilon^{\beta \gamma}_\rho h^{\mu \nu \lambda ) \rho} ,\label{spin6} \\
	\delta_{1}h^{\mu\nu\lambda\rho}=\varepsilon^{\alpha \beta \gamma}\partial_\alpha \partial_\beta \partial_\gamma h^{\mu \nu \lambda \rho}
	+ \partial^{(\mu } \varepsilon^{|\alpha\beta |}_\gamma \partial_\alpha  \partial_\beta h^{\nu \lambda \rho )\gamma }
	+\partial^{(\mu} \partial^\nu\varepsilon^{|\alpha|}_{\beta\gamma} \partial_\alpha h^{ \lambda \rho) \beta\gamma}\nonumber\\+\partial^{(\mu} \partial^\nu \partial^\lambda \varepsilon_{\alpha\beta\gamma} h^{\rho )\alpha\beta\gamma} ,\label{spin4}\\
	\delta_{1}h_{(2)}^{\beta \gamma} =\partial_\mu \partial_\nu \partial_\lambda \partial_\rho  \varepsilon^{\alpha \beta \gamma} \partial_\alpha  h^{\mu \nu \lambda \rho} .\label{spin2}
	\end{gather}
	So we prove that Noether's procedure in this particular case can be done for the construction of the local quartic interaction of the scalar and higher spin fields restricted by transverse and traceless gauge conditions (\ref{2.7}) and (\ref{2.8}).
	
	\section{ Commutator of $\delta_{1}$ transformations for spin four}
	\setcounter{equation}{0}
	In this section we investigate algebra of linear in gauge field transformation (\ref{spin4}) obtained from Noether's proce4dure in previous section:
	\begin{gather}
	\delta_{1}^{(\varepsilon)}h_{\mu\nu\lambda\rho}=\varepsilon^{\alpha \beta \gamma}\partial_\alpha \partial_\beta \partial_\gamma h_{\mu \nu \lambda \rho}
	+ \partial_{(\mu } \varepsilon^{\alpha\beta\gamma} \partial_{|\alpha } \partial_{\beta} h_{\gamma|\nu \lambda \rho ) }+\partial_{(\mu} \partial_\nu\varepsilon^{\alpha\beta\gamma} \partial_{|\alpha} h_{ \beta\gamma|\lambda \rho)}
	\nonumber\\+\partial_{(\mu} \partial_\nu \partial_\lambda \varepsilon^{\alpha\beta\gamma} h_{\rho )\alpha\beta\gamma} .\label{6.1}
	\end{gather}
	The structure of this expression is similar to linear transformation obtained in  \cite{Manvelyan:2010jf} where nonlinear curvature for general higher spin and in particular for spin three case is considered.
First of all for understanding of corresponding gauge algebra we can derive commutator of this linear $\delta_{1}$ transformation (\ref{6.1}) with zero order gauge transformation $\delta_{0}$ (\ref{3.3}) from which our investigation of Noether equation (\ref{3.4}) began. Straightforward calculations leads to the following expression:
\begin{align}
 &\big[\delta_{0}^{(\omega)}\delta^{(\varepsilon)}_{1}-\delta_{0}^{(\varepsilon)}\delta^{(\omega)}_{1}\big]h_{\mu\nu\lambda\rho}
=\partial_{(\mu}\big[\varepsilon^{\alpha\beta\gamma}\partial_{|\alpha}\partial_{\beta}\partial_{\gamma|}\omega_{\nu\lambda\rho)}
 +t_{\nu\lambda\rho)}(\varepsilon,\omega)-(\varepsilon \leftrightarrow \omega)\big] ,\label{com0}\\
 &t_{\nu\lambda\rho}(\varepsilon,\omega)= \partial_{(\nu}\varepsilon^{\alpha\beta\gamma}\partial_{|\alpha}\partial_{\beta|}\omega_{\lambda\rho)\gamma}
 +\partial_{(\nu}\partial_{\lambda}\varepsilon^{\alpha\beta\gamma}\partial_{|\alpha|}\omega_{\rho)\beta\gamma}
 +\frac{1}{3}\partial_{(\nu}\partial_{\lambda}\varepsilon^{\alpha\beta\gamma}\partial_{\rho)}\omega_{\alpha\beta\gamma} .\label{tsmall}
\end{align}
Here we should make two important comments:

First, we see that in (\ref{6.1}) the form of the last three terms is ambiguously defined due to the freedom in the definition of the $\delta_{1}$. This transformation can be modified by adding zero-order (full gradient) transformation with field-dependent parameter. Ruffly speaking we can add $\delta_{0}$ transformation with linear on gauge field parameter to (\ref{6.1}) modifying the last three terms and getting corresponding modification for tensor $t_{\nu\lambda\rho}(\varepsilon,\omega)$ in definition of the commutator (\ref{com0}).

Second following the ideas of \cite{Manvelyan:2010jf} and extracting the same type $\delta_{0}$ terms described above we can rewrite (\ref{6.1}) in the following form:
\begin{align}
\delta_{1}^{(\varepsilon)}h_{\mu\nu\lambda\rho}&= \varepsilon^{\alpha\beta\gamma}\Gamma^{(3)}_{\alpha\beta\gamma;\mu\nu\lambda\rho}(h)
+\partial_{(\mu}\Lambda_{\nu\lambda\rho)}(\varepsilon,h) ,\label{6.2} \\
\Lambda_{\nu\lambda\rho}(\varepsilon,h) &= \varepsilon^{\alpha\beta\gamma}\partial_{\alpha}\partial_{\beta}h_{\gamma\nu\lambda\rho}+
\frac{1}{2}\left[\partial_{(\nu}\varepsilon^{\alpha\beta\gamma}
\partial_{|\alpha}h_{\beta\gamma|\lambda\rho)}-\varepsilon^{\alpha\beta\gamma}
\partial_{(\nu}\partial_{|\alpha}h_{\beta\gamma|\lambda\rho)}\right]\nonumber\\
&+\frac{1}{3}\left[\partial_{(\nu}\partial_{\lambda}\varepsilon^{\alpha\beta\gamma}
h_{\rho)\alpha\beta\gamma}+\varepsilon^{\alpha\beta\gamma}
\partial_{(\nu}\partial_{\lambda}h_{\rho)\alpha\beta\gamma}-\frac{1}{2}
\partial_{(\nu}\varepsilon^{\alpha\beta\gamma}\partial_{\lambda}
h_{\rho)\alpha\beta\gamma}\right] ,\label{6.3}
\end{align}
where\footnote{We use $(\dots)$ and $<\dots>$ brackets for symmetrization of the different set of indices, reserving $[\dots]$ and $\Big[\dots\Big]$ for antisymmetrization.  All other necessary for our case formulas connected with this hierarchy can be found in Appendix B.}
\begin{align}\label{6.4}
\Gamma^{(3)}_{\alpha\beta\gamma;\mu\nu\lambda\rho}(h)&= \partial_{\alpha}\partial_{\beta}\partial_{\gamma}
h_{\mu\nu\lambda\rho}-\frac{1}{3}\partial_{<\alpha}\partial_{\beta}\partial_{(\mu}
h_{\nu\lambda\rho)\gamma>}+\frac{1}{3}\partial_{<\alpha}\partial_{(\mu}\partial_{\nu}
h_{\lambda\rho)\beta\gamma>}\nonumber\\
&-\partial_{(\mu}\partial_{\nu}\partial_{\lambda}
h_{\rho)\alpha\beta\gamma} ,
\end{align}
is the third for spin four gauge field (last before Curvature) Christoffel Symbol  in deWit-Freedman hierarchy  of connections defined in \cite{deWit:1979sib}\footnote{It is worth noting here that the transformation of the additional spin 6 gauge field through the spin four gauge field (\ref{spin6}) can also be written in a similar to (\ref{6.2}) form:
	\begin{equation*}
	\delta_1 h^{(6)}_{\mu \nu \lambda \alpha \beta \gamma}=\varepsilon^{\rho}_{(\alpha \beta}\Gamma^{(1)}_{|\rho|;\gamma \mu \nu \lambda)}(h) +\partial_{(\alpha}[\varepsilon^{\rho}_{\beta \gamma} h_{\mu \nu \lambda )\rho}],
	\end{equation*}
	where $\Gamma^{(1)}_{\rho;\gamma \mu \nu \lambda)}(h)$ is first generalized Christoffel symbol for spin four field defined in (\ref{7.2}).}.	
The key point of the splitting (\ref{6.2}) is the simple form of zero order on field gauge transformation of connection (\ref{6.4}) (see formulas (\ref{7.2})-(\ref{7.9}) for details):
\begin{equation}\label{6.5}
\delta^{(\varepsilon)}_{0}\Gamma^{(3)}_{\alpha\beta\gamma;\mu\nu\lambda\rho}(h)=
-4\partial_{\mu}\partial_{\nu}\partial_{\lambda}\partial_{\rho}\varepsilon_{\alpha\beta\gamma} ,
\end{equation}
and possibility in the future calculations \emph{to identify in r.h.s of commutator different symmetries by existence of the terms in the form of Christoffel symbols or Generalized Curvatures (in some case with symmetrized derivatives) contracted with composite parameters  like in (\ref{6.2}) but with different rank and symmetry structure of the indices for composite parameters. So from now on  we call such a type of terms as a "regular" }.   	
In this way we see that expressions (\ref{6.1})-(\ref{6.5}) is really looks like higher spin generalization of the gauge transformation (\ref{1.16}) (Lie derivative) and usual Christoffel symbol for linearized gravity\footnote{Note that most common definition of Christoffel symbol $\Gamma^{\beta}_{\mu\nu}(g)
	=\frac{1}{2}g^{\beta\alpha}(\partial_{(\mu}g_{\nu)\alpha}
	-\partial_{\alpha}g_{\mu\nu})$ for general metric $g_{\mu\nu}$ relates with our definition after linearization in the flat background in the following way $\Gamma^{\beta}_{\mu\nu}(\eta_{\mu\nu}+h_{\mu\nu})=-\frac{1}{2}\eta^{\beta\alpha}\Gamma^{(1)}_{\alpha;\mu\nu}(h)$.}
\begin{align}
\delta_{1}^{(\varepsilon)}h_{\mu\nu} &= \mathfrak{L}_{\varepsilon^{\lambda}}h_{\mu\nu}=\varepsilon^{\alpha}
\Gamma^{(1)}_{\alpha;\mu\nu}+\partial_{(\mu}\left(\varepsilon^{\alpha}h_{\nu)\alpha}\right) ,
\label{6.6}\\
\Gamma^{(1)}_{\alpha;\mu\nu} &=\partial_{\alpha}h_{\mu\nu} -\partial_{(\mu}h_{\nu)\alpha} ,\label{6.7}\\
\delta^{(\varepsilon)}_{0}\Gamma^{(1)}_{\alpha;\mu\nu}(h)&=
-2\partial_{\mu}\partial_{\nu}
\varepsilon_{\alpha} .\label{6.8}
\end{align}
Using representation (\ref{6.2}) and transformation rule (\ref{6.5}) we can derive the following expression for commutator:
\begin{align}\label{6.9}
[\delta^{(\omega)}_{1},\delta^{(\varepsilon)}_{1}]h_{\mu\nu\lambda\rho}&= \varepsilon^{\alpha\beta\gamma}\Gamma^{(3)}_{\alpha\beta\gamma;\mu\nu\lambda\rho}
(\delta^{(\omega)}_{1}h)-4\varepsilon^{\alpha\beta\gamma}\partial_{\mu}
\partial_{\nu}\partial_{\lambda}\partial_{\rho}\Lambda_{\alpha\beta\gamma}(\omega,h)\nonumber\\
&+ \partial_{(\mu}\Lambda_{\nu\lambda\rho)}(\varepsilon,\delta^{(\omega)}_{1}h)-(\varepsilon \leftrightarrow \omega) .
\end{align}
Then taking int account that all symmetrized full gradients in r.h.s we can drop as a trivial $\delta_{0}$  contribution from composite symmetric third rank gauge parameter linear in gauge field, we can first of all drop second line in (\ref{6.9}). Then we can put four $\mu,\nu,\lambda,\rho,$ derivatives  in second term of first line from $\Lambda_{\alpha\beta\gamma}$ to parameter $\varepsilon^{\alpha\beta\gamma}$ and integrate using formula (\ref{6.5}) and came to the following expression
\begin{align}\label{6.10}
[\delta^{(\omega)}_{1},\delta^{(\varepsilon)}_{1}]h_{\mu\nu\lambda\rho}&\sim \varepsilon^{\alpha\beta\gamma}\Gamma^{(3)}_{\alpha\beta\gamma;\mu\nu\lambda\rho}
(\delta^{(\omega)}_{1}h)+\Gamma^{(3)}_{\alpha\beta\gamma;\mu\nu\lambda\rho}(h)
\delta^{(\omega)}_{0}\Lambda^{\alpha\beta\gamma}(\varepsilon,h)-(\varepsilon \leftrightarrow \omega) ,
\end{align}
where $\sim$ means an equality up to any $\delta_{0}$ variations with composed field dependent parameter described above or delta zero variation with usual parameter $\varepsilon$ or $\omega$ from any second order on gauge field tensor.
At this point it is worth  to note that considering perturbative  on linearized gauge field deformation of the initial gauge transformation regulated by Noether's procedure
\begin{equation}\label{6.11}
	\delta^{(\epsilon)}h_{\mu\nu\lambda\rho}
	=(\delta^{(\epsilon)}_{0}+\delta^{(\epsilon)}_{1}
	+\delta^{(\epsilon)}_{2}+\dots)h_{\mu\nu\lambda\rho},
	\end{equation}
	for commutator on the linear level on gauge field we obtain:
	\begin{equation}\label{6.12}
	\Big\{[\delta^{(\omega)}, \delta^{(\epsilon)}]h_{\mu\nu\lambda\rho}\Big\}_{1}=([\delta^{(\omega)}_{1}, \delta^{(\epsilon)}_{1}]+\delta^{(\omega)}_{0}\delta^{(\varepsilon)}_{2}-\delta^{(\varepsilon)}_{0}\delta^{(\omega)}_{2})h_{\mu\nu\lambda\rho} .
	\end{equation}
	
	So we see that we can factorize in right hand side of our commutator of the first order gauge transformation two type of trivial terms:
	\begin{itemize}
		\item Symmetrized full derivatives from composed gauge parameter linear in gauge fields $\partial_{(\mu}\tilde{\Lambda}_{\nu\lambda\rho)}(\varepsilon,\omega,h)-(\varepsilon \leftrightarrow \omega)$.
		\item The terms which  can be classified  as a second part of r.h.s of (\ref{6.12}):\newline  $\delta^{(\omega)}_{0}\delta^{(\varepsilon)}_{2}h_{\mu\nu\lambda\rho}-(\varepsilon
		\leftrightarrow \omega)$,  and we can throw them out also to understand algebra of two $\delta_{1}$ transformations.
	\end{itemize}
	
	Now following this simple methodology we can present final result for commutator hiding long and tedious calculation in Appendix C:
	\begin{eqnarray}
	&&[\delta^{(\omega)}_{1},\delta^{(\varepsilon)}_{1}]h_{\mu\nu\lambda\rho}\sim  \left[\varepsilon^{\delta\sigma\eta}\partial_{\delta}\partial_{\sigma}\partial_{\eta}
	\omega^{\alpha\beta\gamma}+T^{\alpha\beta\gamma}(\partial,\varepsilon,\omega)
	\right] \Gamma^{(3)}_{\alpha\beta\gamma;\mu\nu\lambda\rho}(h)\nonumber\\
	&&+3\varepsilon^{\delta\sigma\eta}\partial_{\delta}
	\partial_{\sigma}\omega^{\alpha\beta\gamma}
	R^{(4)}_{\eta\alpha\beta\gamma;\mu\nu\lambda\rho}(h)+\frac{9}{20}\varepsilon_{\delta}^{\sigma\eta}
	\partial^{[\delta}\omega^{\alpha]\beta\gamma}
	\partial_{(\sigma}R^{(4)}_{\eta\alpha\beta\gamma);\mu\nu\lambda\rho}(h)\nonumber\\
	&&+[Rem]_{\mu\nu\lambda\rho}(\varepsilon,\omega,h)-(\varepsilon\leftrightarrow\omega) \,,\label{6.13}
	\end{eqnarray}
	
	where:
	\begin{eqnarray}
	T^{\alpha\beta\gamma}(\partial,\varepsilon,\omega)&=&\frac{1}{4}
	\partial^{(\alpha}\partial^{\beta}
	\varepsilon^{\delta\sigma\eta}\delta_{0}^{(\omega)}
	h_{\,\,\delta\sigma\eta}^{\gamma)}-\frac{5}{48}\partial^{(\alpha}\varepsilon^{\delta\sigma\eta}
	\partial^{\beta}\delta_{0}^{(\omega)}h^{\gamma)}_{\delta\sigma\eta} +\frac{7}{16}\partial^{(\alpha}\varepsilon^{\delta\sigma\eta}
	\partial_{\delta}\delta_{0}^{(\omega)}h_{\sigma\eta}^{\,\,\,\beta\gamma)}
	\nonumber\\
	&-&\frac{1}{16}\partial^{\delta}\varepsilon^{\sigma\eta(\alpha}
	\partial^{\beta}\delta_{0}^{(\omega)}h^{\gamma)}_{\delta\sigma\eta}+\frac{1}{16}\partial^{\delta}\varepsilon^{\sigma\eta(\alpha}
	\partial_{\delta}\delta_{0}^{(\omega)}h^{\beta\gamma)}_{\sigma\eta}\,,\nonumber\\\label{6.14}
	\end{eqnarray}
	and
	\begin{align}
	&[Rem]_{\mu\nu\lambda\rho}(\varepsilon,\omega,h)=\nonumber\\
	&\frac{9}{20}\varepsilon_{\delta}^{\eta\sigma}
	\partial^{[\delta}\omega^{\alpha]\beta\gamma}
	\partial_{(\mu}R^{(3)}_{\nu\lambda\rho);\eta\beta\gamma}(H^{(3)}_{[\alpha\sigma]})
	+\frac{3}{2}\partial_{(\mu}\varepsilon_{\delta}^{\eta\sigma}
	\partial^{[\delta}\omega^{\alpha]\beta\gamma} R^{(3)}_{\nu\lambda\rho);\eta\beta\gamma}(H^{(3)}_{[\alpha\sigma]})
	-\frac{9}{40}\varepsilon_{\delta}^{\eta\sigma}
	\partial^{[\delta}\omega^{\alpha]\beta\gamma}\partial_{(\mu}R^{(3)}_{\nu\lambda\rho);\eta\alpha\gamma}(H^{(3)}_{[\beta\sigma]})
	\nonumber\\
	&+\frac{3}{8}\varepsilon^{\eta}_{\sigma\delta}\partial^{[\sigma}
	\partial^{\big[\delta}\omega^{\alpha\big]\beta]\gamma}\partial_{(\mu}
	\Gamma^{(2)}_{\beta\gamma;\nu\lambda\rho)}(H^{(3)}_{[\eta\alpha]})
	+\frac{1}{2}\partial_{(\mu}
	\varepsilon^{\eta}_{\delta\sigma}\partial^{[\sigma}
	\partial^{\big[\delta}\omega^{\alpha\big]\beta]\gamma}
	\Gamma^{(2)}_{\beta\gamma;\nu\lambda\rho)}(H^{(3)}_{[\eta\alpha]})
	\quad\qquad\nonumber\\
	&+\frac{3}{8}\varepsilon^{\sigma\eta}_{\delta}
	\partial^{[\delta}\omega^{\alpha]\beta\gamma}\partial_{(\mu}\partial_{\nu}
	\Gamma^{(1)}_{\gamma;\lambda\rho)}(H^{(2)}_{[\eta\alpha][\sigma\beta]})+
	\frac{1}{2}\partial_{(\mu}\varepsilon^{\sigma\eta}_{\delta}
	\partial^{[\delta}\omega^{\alpha]\beta\gamma}
	\partial_{\nu}\Gamma^{(1)}_{\gamma;\lambda\rho)}(H^{(2)}_{[\eta\alpha][\sigma\beta]})\nonumber\\&
	+\frac{3}{4}\partial_{(\mu}\partial_{\nu}
	\varepsilon^{\sigma\eta}_{\delta}
	\partial^{[\delta}\omega^{\alpha]\beta\gamma}
	\Gamma^{(1)}_{\gamma;\lambda\rho)}(H^{(2)}_{[\eta\alpha][\sigma\beta]})\label{6.15}
	\end{align}
	is remaining part of commutator contained transformation described by composed gauge parameter with mixed symmetry of indices in the form of one or two antisymmetrized pairs.

To be more precise when classifying terms on the right side of (\ref{6.13}) let us consider each line separately:
\begin{enumerate}
  \item The first line describes spin four gauge transformation with composite \emph{symmetric rank 3 tensor parameter} in the form
  \begin{equation}\label{reg1}
  [\omega,\varepsilon]^{\alpha\beta\gamma}\Gamma^{(3)}_{\alpha\beta\gamma;\mu\nu\lambda\rho}(h) ,
\end{equation}
where
  \begin{eqnarray}
    && [\omega,\varepsilon]^{\alpha\beta\gamma}=\varepsilon^{\delta\sigma\eta}\partial_{\delta}\partial_{\sigma}\partial_{\eta}
	\omega^{\alpha\beta\gamma}+T^{\alpha\beta\gamma}(\partial,\varepsilon,\omega)-(\varepsilon\leftrightarrow\omega). \label{omal}
  \end{eqnarray}
  \item The second line also corresponds to the transformation of the spin four gauge field in respect to gauge transformation with symmetric tensor parameter. But in this case we have \emph{symmetric tensor parameters of rank 4 and 5}, which means that it is transformation coming from gauge field with spin 5 and 6 and our spin four gauge field participates in these transformations through the spin four gauge invariant (in zero order on field transformations) curvature. In other words we have here  regular terms in the form
      \begin{eqnarray}
      &&\Omega_{(4)}^{\eta\alpha\beta\gamma\delta}(\varepsilon,\omega)R^{(4)}_{\eta\alpha\beta\gamma;\mu\nu\lambda\rho}(h) , \label{320} \\
      &&\Omega_{(5)}^{\sigma\eta\alpha\beta\gamma\delta}(\varepsilon,\omega)\partial_{(\sigma}
      R^{(4)}_{\eta\alpha\beta\gamma);\mu\nu\lambda\rho}(h) ,\label{321}
      \end{eqnarray}
      where
      \begin{eqnarray}
        &&\Omega_{(4)}^{\eta\alpha\beta\gamma\delta}(\varepsilon,\omega)=\frac{3}{4}\varepsilon^{\delta\sigma(\eta}\partial_{\delta}
	\partial_{\sigma}\omega^{\alpha\beta\gamma)}-(\varepsilon\leftrightarrow\omega) ,\label{322} \\
        && \Omega_{(5)}^{\sigma\eta\alpha\beta\gamma\delta}(\varepsilon, \omega)=\frac{9}{200}\varepsilon^{\delta(\sigma\eta}
	\partial_{\delta}\omega^{\alpha\beta\gamma)}-\frac{3}{200}\varepsilon_{\delta}^{(\sigma\eta}
	\partial^{\alpha}\omega^{\beta\gamma)\delta}-(\varepsilon\leftrightarrow\omega) .\qquad\label{323}
      \end{eqnarray}
  \item Now we analyze the third line of (\ref{6.13}) or eight terms in expression (\ref{6.15}). First of all we see that in this remaining part of commutator our spin four field expressed through the reduced curvatures and Christoffel symbols defined in Appendix B (see (\ref{7.20})-(\ref{7.28}) and (\ref{7.35}), (\ref{7.36})). All such a objects possess one (first two lines of (\ref{6.15})) ore two (remaining two lines of (\ref{6.15})) pair of antisymmetrized indices contracted with composed gauge parameter. Therefore they could describe some mixed symmetry field gauge transformation acting on spin four symmetric gauge field. For example first term in (\ref{6.15}) we can rewrite in the form:
      \begin{eqnarray}
        &&\Omega^{[\alpha\sigma],\eta\beta\gamma}_{[2],(3)}
        \partial_{(\mu}R^{(3)}_{\nu\lambda\rho);\eta\beta\gamma}(H^{(3)}_{[\alpha\sigma]})\label{324}
      \end{eqnarray}
      where
      \begin{eqnarray}
        &&\Omega^{[\alpha\sigma],\eta\beta\gamma}_{[2],(3)} =\frac{3}{40}(\varepsilon^{\delta(\eta[\sigma}
	\partial_{\delta}\omega^{\alpha]\beta\gamma)}- \varepsilon_{\delta}^{(\eta[\sigma}
	\partial^{\alpha]}\omega^{\beta\gamma)\delta}) \label{325}
      \end{eqnarray}
      and in the same way the sixth term with two pair of antisymmetrized  indices we can express as
      \begin{equation}\label{326}
       \Omega^{[\eta\alpha],[\sigma\beta],\gamma}_{[2],[2],(1)} \partial_{(\mu}\partial_{\nu}
	\Gamma^{(1)}_{\gamma;\lambda\rho)}(H^{(2)}_{[\eta\alpha][\sigma\beta]})
      \end{equation}
      where composit parameter is
      \begin{equation}\label{327}
        \Omega^{[\eta\alpha],[\sigma\beta],\gamma}_{[2],[2],(1)}= \frac{3}{32}(\varepsilon^{\delta\big[\sigma[\eta}
	\partial_{\delta}\omega^{\alpha]\beta\big]\gamma}-\varepsilon^{\big[\sigma[\eta}_{\delta}
	\partial^{\alpha]}\omega^{\beta\big]\delta\gamma})
      \end{equation}
This type of terms (first, third, fourth and sixth in (\ref{6.15})) with mixed symmetry composed parameters we can still call "regular". But four remaining terms of (\ref{6.15}) (second, fifth, seventh and eighth) we cannot transform to regular form because they all have non contracted derivatives from one (non composed) gauge parameter ($\partial_{\mu}\varepsilon$ or $\partial_{\mu}\partial_{\nu}\varepsilon$) and we call these terms irregular because do not have at the moment interpretation of them in means of additional symmetries or equation of motion of theory under construction. But at least we can clime that all irregular terms are in the mixed symmetry parameter sector.
\end{enumerate}

So we see that our \emph{commutator of spin four linear on gauge field transformations produce regular terms coming from gauge transformation of symmetric tensors with spin $s<6$} and remaining irregular transformation with mixed symmetry gauge field parameters.
	\section{Conclusion}
	In this paper, we considered the possibility to construct local quartic interaction in the special case of two higher spin gauge fields and two scalars.
	Restricting ourselves to the traceless and transverse physical gauge for spin four HS fields we obtained a solution of the Noether's equation adding cubic interaction with additional spin six field in the form of gauge field-current (\ref{3.9}). As an additional bonus of this construction, we derived linear on-field gauge transformation for spin four filed (\ref{spin4}) getting a possibility to investigate the corresponding commutator. Performing complicated calculations and using the formalism of generalized Christoffel symbols \cite{deWit:1979sib} we classified the right-hand side of commutator and understood that general algebra of $\delta^{(\varepsilon)}_{1}$ transformation closes on not only spin four gauge transformation. The r.h.s. of commutator includes the sector of transformations of spin four field in respect to gauge transformations with parameters corresponding to spin four and higher up to spin six gauge fields. All other terms include only transformations with parameters coming from transformations of additional fields with mixed symmetry of indices including one or two antisymmetric pairs (\ref{6.15}). Investigation and analyses of this mixed symmetry sector of r.h.s of commutator can be done in future in separate work.
	
	\section*{Acknowledgments}
	This work was supported by the Science Committee of RA, in the frames of the research project \# 21AG-1C060. We are grateful to Karapet Mkrtchyan for useful comments and suggestions.

	\section*{Appendix A: Details of the Noether's procedure}
	\renewcommand{\theequation}{A.\arabic{equation}}\setcounter{equation}{0}
	\subsection*{Tuning of cubic interactions}
	To prove (\ref{5.7last})-(\ref{spin2}) we should first derive special relations. After some partial integration and symmetrization of indices we can prove for our constrained field and parameters (\ref{2.7})-(\ref{2.10}) the following relations
	\begin{eqnarray}
	&&\frac{1}{15}\left[\partial_\mu \varepsilon^{(\alpha \beta \gamma} h^{\nu \lambda \rho )\mu}-\varepsilon^{\mu (\alpha\beta} \partial_\mu h^{\gamma \nu \lambda \rho)}\right]\partial_{(\nu}\partial_{\lambda} J^{(4)}_{\rho \alpha \beta \gamma)}=\nonumber\\
	&& -\left[\partial_\alpha   \varepsilon^{\mu\nu (\beta} \partial_\mu  \partial_\nu h^{\gamma \lambda \rho ) \alpha } -\partial_\mu \partial_\nu \varepsilon^{\alpha ( \beta \gamma} \partial_\alpha h^{ \lambda \rho ) \mu \nu}\right]  J^{(4)}_{\lambda \rho \beta \gamma}\nonumber\\
	&&+\left[\partial_\alpha \partial_\beta \partial_\gamma \varepsilon^{(\mu\nu\lambda} h^{\rho ) \alpha\beta\gamma }-\varepsilon^{\alpha \beta \gamma} \partial_\alpha \partial_\beta \partial_\gamma  h^{\mu \nu \lambda \rho}\right] J^{(4)}_{\mu \nu \lambda \rho} ,\label{4.3}
	\end{eqnarray}
	
	\begin{eqnarray}
	&&\frac{1}{30}\left[\partial_\mu \varepsilon^{(\alpha \beta \gamma} h^{\nu \lambda \rho )\mu}-\varepsilon^{\mu (\alpha\beta} \partial_\mu h^{\gamma \nu \lambda \rho)}\right]\partial_{(\nu}\partial_{\lambda}\partial_\rho \partial_\alpha J^{(2)}_{\beta \gamma)}=\nonumber\\
	&+&[\partial_\mu \partial_\nu \partial_\gamma \varepsilon^{\alpha \beta ( \lambda} \partial_\alpha \partial_\beta h^{\rho )\mu \nu \gamma }
	-\partial_\mu \partial_\nu \varepsilon^{\alpha \beta \gamma} \partial_\alpha \partial_\beta \partial_\gamma h^{\mu \nu \lambda \rho}] J^{(2)}_{\lambda\rho}\nonumber\\
	&+&\frac{3}{2}\partial_\mu \partial_\nu \partial_\lambda \partial_\rho  \varepsilon^{\alpha \beta \gamma} \partial_\alpha  h^{\mu \nu \lambda \rho} J^{(2)}_{\beta \gamma} .\quad\quad\quad \label{4.4}
	\end{eqnarray}
	Using relation (\ref{4.3}) we can bring improvement for discrepancy in numbers a front of $J^{(4)}$ terms in (\ref{3.5}). The second relation (\ref{4.4}) is suitable for cancellation of the fifth line of (\ref{3.5}) connected with $J^{(2)}$ current.
	Finally we obtain instead of (\ref{3.5}) the following relation:
	\begin{eqnarray}
	\delta_1 (h^{\mu \nu \lambda \rho} \partial_\mu \partial_\nu \Phi \partial_\lambda \partial_\rho \Phi)&=&\frac{1}{3}\delta_1 (h^{\mu \nu \lambda \rho}J^{(4)}_{\mu\nu\lambda\rho})=\delta_1 h^{\mu \nu \lambda \rho} \partial_\mu \partial_\nu \Phi \partial_\lambda \partial_\rho \Phi\nonumber\\
	&+&\frac{1}{50}\left[\varepsilon^{\mu (\alpha\beta} \partial_\mu h^{\gamma \nu \lambda \rho)}-\partial_\mu \varepsilon^{(\alpha \beta \gamma} h^{\nu \lambda \rho )\mu}\right] \tilde{J}^{(6)}_{\nu \lambda \rho \alpha \beta \gamma}\nonumber\\
	&+& \frac{1}{6} \left[\partial_\alpha   \varepsilon^{\mu\nu (\beta} \partial_\mu  \partial_\nu h^{\gamma \lambda \rho ) \alpha } -\partial_\mu \partial_\nu \varepsilon^{\alpha ( \beta \gamma} \partial_\alpha h^{ \lambda \rho ) \mu \nu}\right]  J^{(4)}_{\lambda \rho \beta \gamma}\nonumber\\
	&+&\frac{1}{6}  \left[\partial_\alpha \partial_\beta \partial_\gamma \varepsilon^{(\mu\nu\lambda} h^{\rho ) \alpha\beta\gamma }-\varepsilon^{\alpha \beta \gamma} \partial_\alpha \partial_\beta \partial_\gamma  h^{\mu \nu \lambda \rho}\right] J^{(4)}_{\mu \nu \lambda \rho}\nonumber\\
	&+&\frac{1}{2}\partial_\mu \partial_\nu \partial_\lambda \partial_\rho  \varepsilon^{\alpha \beta \gamma} \partial_\alpha  h^{\mu \nu \lambda \rho} J^{(2)}_{\beta \gamma}\,,\quad\quad\quad \label{4.5}
	\end{eqnarray}
	where modified $\tilde{J}^{(6)}$ is
	\begin{eqnarray}\label{4.6}
	\tilde{J}^{(6)}_{\nu \lambda \rho \alpha \beta \gamma}=J^{(6)}_{\nu \lambda \rho \alpha \beta \gamma}+\frac{1}{9} \partial_{(\alpha} \partial_\beta J^{(4)}_{ \gamma \nu \lambda \rho )}+\frac{1}{3} \partial_{(\nu} \partial_\lambda \partial_\rho \partial_\alpha  J^{(2)}_{\beta \gamma)} .
	\end{eqnarray}
	The last term in expression (\ref{4.5}) also could be canceled due to relation
	\begin{eqnarray}
	&&\left[\partial_\alpha   \varepsilon^{\mu\nu (\beta} \partial_\mu  \partial_\nu h^{\gamma \lambda \rho ) \alpha } -\partial_\mu \partial_\nu \varepsilon^{\alpha ( \beta \gamma} \partial_\alpha h^{ \lambda \rho ) \mu \nu}\right]  \partial_\lambda \partial_\rho J^{(2)}_{\beta \gamma}\nonumber\\
	&+&\left[\partial_\alpha \partial_\beta \partial_\gamma \varepsilon^{(\mu\nu\lambda} h^{\rho ) \alpha\beta\gamma }-\varepsilon^{\alpha \beta \gamma} \partial_\alpha \partial_\beta \partial_\gamma  h^{\mu \nu \lambda \rho}\right] \partial_\mu \partial_\nu J^{(2)}_{\lambda \rho}=\nonumber\\
	&&6\partial_\mu \partial_\nu \partial_\lambda \partial_\rho  \varepsilon^{\alpha \beta \gamma} \partial_\alpha  h^{\mu \nu \lambda \rho} J^{(2)}_{\beta \gamma} .\quad\quad\quad \label{4.7}
	\end{eqnarray}
	But in this case we see that this we can do only after adding to initial $J^{(4)}$ current a symmetrized double gradient of $J^{(2)}$:
	\begin{equation}\label{4.8}
	\tilde{J}^{(4)}_{\nu \lambda \rho \gamma}= J^{(4)}_{\nu \lambda \rho \gamma}+\frac{1}{2}\partial_{(\nu} \partial_\lambda J^{(2)}_{ \rho \gamma )} ,
	\end{equation}
	which is possible but needs special consideration out of our restrictions on spin 4 gauge field. So we prefer to keep initial spin 4 current unchanged and cancel this term by traceless Stueckelberg like transformation of the spin two gauge field from linear coupling with $J^{(2)}$ current:
	\begin{equation}\label{4.9}
	\delta_{1}h_{(2)}^{\beta \gamma} \sim \partial_\mu \partial_\nu \partial_\lambda \partial_\rho  \varepsilon^{\alpha \beta \gamma} \partial_\alpha  h^{\mu \nu \lambda \rho} .
	\end{equation}
	Note that this transformation is always traceless due to tracelessness of gauge parameter in Fronsdal formulation.
	So we need to work out only first four lines of (\ref{4.3})
	\subsection*{Integration and interaction}
	Now we start to integrate expression:
	\begin{eqnarray}
	&&\frac{1}{50}\left[\varepsilon^{\mu (\alpha\beta} \partial_\mu h^{\gamma \nu \lambda \rho)}-\partial_\mu \varepsilon^{(\alpha \beta \gamma} h^{\nu \lambda \rho )\mu}\right] \tilde{J}^{(6)}_{\nu \lambda \rho \alpha \beta \gamma}\nonumber\\
	&+& \frac{1}{6} \left[\partial_\alpha   \varepsilon^{\mu\nu (\beta} \partial_\mu  \partial_\nu h^{\gamma \lambda \rho ) \alpha } -\partial_\mu \partial_\nu \varepsilon^{\alpha ( \beta \gamma} \partial_\alpha h^{ \lambda \rho ) \mu \nu}\right]  J^{(4)}_{\lambda \rho \beta \gamma}\nonumber\\
	&+&\frac{1}{6}  \left[\partial_\alpha \partial_\beta \partial_\gamma \varepsilon^{(\mu\nu\lambda} h^{\rho ) \alpha\beta\gamma }-\varepsilon^{\alpha \beta \gamma} \partial_\alpha \partial_\beta \partial_\gamma  h^{\mu \nu \lambda \rho}\right] J^{(4)}_{\mu \nu \lambda \rho} .\label{5.1}
	\end{eqnarray}
	To extract interactions and linear on gauge field transformations we can use the following important formulas:
	\begin{gather}
	\partial_\mu \varepsilon^{\alpha \beta \gamma}=\delta_0 h^{\alpha \beta \gamma}_{\mu}- \partial^{(\alpha} \varepsilon^{ \beta \gamma )}_\mu , \label{5.2}
	\\
	\partial_\mu \partial_\nu \varepsilon^{\alpha \beta \gamma}=\frac{1}{2}\partial_{(\nu}\delta_0 h^{\alpha \beta \gamma}_{\mu)}-\frac{1}{2}\partial^{(\alpha} \delta_0 h^{\beta \gamma )}_{\mu \nu}+\partial^{(\alpha} \partial^\beta \varepsilon^{ \gamma )}_{\mu \nu} ,\label{5.3}
	\\
	\partial_\mu \partial_\nu \partial_\lambda \varepsilon^{\alpha \beta \gamma}=\frac{1}{3}\partial_{(\nu} \partial_\lambda \delta_0 h^{\alpha \beta \gamma}_{\mu)}-\frac{1}{6}\partial^{(\alpha}\partial_{(\lambda} \delta_0 h^{\beta \gamma )}_{\mu \nu)}+\frac{1}{3} \partial^{(\alpha} \partial^\beta \delta_0 h^{\gamma)}_{\mu \nu \lambda} -\partial^\alpha \partial^\beta \partial^\gamma \varepsilon_{\mu \nu \lambda} .\label{5.4}
	\end{gather}
	Using (\ref{5.2}) we can immediately integrate first line of (\ref{5.1}) and obtain interaction Lagrangian:
	\begin{equation}\label{5.5}
	L^{1}_{2}=\frac{1}{10} h^{\alpha \beta \gamma}_\mu h^{\nu \lambda \rho \mu}\tilde{J}^{(6)}_{\nu \lambda \rho \alpha \beta \gamma} .
	\end{equation}
	The remaining part of first line of (\ref{5.1}) can be removed by gauge transformation of the spin 6 field
	\begin{gather}
	\delta_1 h_{(6)}^{\mu \nu \lambda \alpha \beta \gamma}=\varepsilon^{\rho (\alpha \beta}\partial_{\rho}h^{\gamma \mu \nu \lambda )}+\partial^{(\alpha}\varepsilon^{\beta \gamma}_\rho h^{\mu \nu \lambda ) \rho} ,\label{5.6}
	\end{gather}
	in cubic part (\ref{3.9}).
	Then we turn to the second and third line of (\ref{5.1}). Applying (\ref{5.1}) and (\ref{5.3}) to the first terms of second and third lines of (\ref{5.1}) we see immediately that for integration of two terms $\delta_{0}h\partial\partial h + \partial\partial\delta_{0}h h=\delta_{0}(h\partial\partial h)$ coming from different brackets we need the same coefficient a front of second and third line of (\ref{5.1}) and corresponding identity (\ref{4.3}) helped us to improve this discrepancy in (\ref{3.5}). Finally using (\ref{5.3}) for other terms we arrive to the following interaction terms coming from these two lines:
	\begin{gather}
	L^{2}_{2}= - \frac{2}{3} h^{\alpha \beta \gamma}_{\mu} \partial_\alpha  \partial_\beta h^{\mu\nu\lambda\rho}J^{(4)}_{\nu \lambda \rho \gamma}
	+\frac{1}{2} \partial_{\nu} h^{\alpha  \beta \gamma}_{\mu} \partial_\alpha h^{ \mu\nu\lambda\rho} J^{(4)}_{\lambda \rho \beta \gamma}
	-\frac{1}{4} \partial^{\alpha}  h^{\beta \gamma }_{\mu \nu} \partial_\alpha h^{\mu\nu\lambda\rho} J^{(4)}_{\lambda \rho \beta \gamma}\nonumber \\
	- \partial^{\beta}  h^{\alpha \gamma }_{\mu \nu} \partial_\alpha h^{\mu\nu\lambda\rho}J^{(4)}_{\lambda \rho \beta \gamma}
	+\frac{1}{3} \partial^\beta  h^{\gamma}_{\mu \nu \lambda} \partial^{\alpha}  h^{\mu\nu\lambda\rho}J^{(4)}_{\rho \alpha \beta \gamma}.\label{5.7}
	\end{gather}
	From remainder we can extract $\delta_{1}h^{\mu\nu\lambda\rho}$
	\begin{gather}
	\delta_{1}h^{\mu\nu\lambda\rho}\sim \nonumber\\\varepsilon^{\alpha \beta \gamma}\partial_\alpha \partial_\beta \partial_\gamma h^{\mu \nu \lambda \rho}
	+ \partial^{(\mu } \varepsilon^{|\alpha\beta |}_\gamma \partial_\alpha  \partial_\beta h^{\nu \lambda \rho )\gamma }
	+\partial^{(\mu} \partial^\nu\varepsilon^{|\alpha|}_{\beta\gamma} \partial_\alpha h^{ \lambda \rho) \beta\gamma}+\partial^{(\mu} \partial^\nu \partial^\lambda \varepsilon_{\alpha\beta\gamma} h^{\rho )\alpha\beta\gamma} .\label{5.8}
	\end{gather}
	
	After all this manipulation we still have four remaining terms of two types:
	
	First two remaining terms contain divergences of spin 4 current:
	\begin{gather}
	\partial_{\lambda} \partial^{( \nu} \varepsilon^{\beta \gamma \mu )}  h^{\rho \lambda }_{\mu \nu}\partial^{\alpha} J^{(4)}_{\rho \alpha \beta \gamma}
	-\frac{2}{3} \partial^\beta \partial^{(\gamma} \varepsilon^{\mu \nu \lambda)} h^{\rho }_{\mu \nu \lambda }\partial^{\alpha}  J^{(4)}_{\rho \alpha \beta \gamma}=
	\nonumber  \\
	-\frac{1}{6}\partial^{(\mu}\big(\partial_\beta \partial_{\gamma}\varepsilon^{\nu \lambda}_\alpha h^{\rho ) \alpha \beta \gamma}\big) J^{(4)}_{\mu \nu \lambda \rho }
	+\frac{1}{18}\partial^{(\mu}\big(\partial^\nu \partial^\lambda \varepsilon_{\alpha \beta \gamma} h^{\rho) \alpha \beta \gamma} \big)J^{(4)}_{\mu \nu \lambda \rho } .\label{5.9}
	\end{gather}
	We can cancel them introducing first order deformation of $\delta_{0}h^{\mu\nu\lambda\rho}$
	\begin{gather}
	\bar{\delta}_0 h^{\mu \nu \lambda \rho}\sim \partial^{(\mu}\big(\partial_\beta \partial_{\gamma}\varepsilon^{\nu \lambda}_\alpha h^{\rho ) \alpha \beta \gamma}\big) -\frac{1}{3}\partial^{(\mu}\big(\partial^\nu \partial^\lambda \varepsilon_{\alpha \beta \gamma} h^{\rho) \alpha \beta \gamma} \big) .\label{5.10}
	\end{gather}

	Second two remainders contain contractions between derivatives of gauge parameter and gauge fields:
	\begin{gather}
	-\frac{4}{3} \partial^{\alpha} \varepsilon^{ \beta \gamma }_\mu \partial_\alpha  \partial_\beta h^{\nu \lambda \rho  \mu }J^{(4)}_{\nu \lambda \rho \gamma}
	- 2\partial^{\alpha} \partial^\beta \varepsilon^{ \gamma }_{\mu \nu} \partial_\alpha h^{ \lambda \rho  \mu \nu} J^{(4)}_{\lambda \rho \beta \gamma} .\label{5.11}
	\end{gather}
	Using the following (up to total derivatives) identity:
	\begin{gather}
	\partial_{\mu} A \partial^{\mu} B C = \frac{1}{2} (A B \Box C - \Box A B C - A \Box B C ) ,\label{5.12}
	\end{gather}
	and taking into account that for our traceless and transversal gauge field on-shell condition means just $\Box h^{\mu \nu \lambda \rho}=0$ and our gauge parameter is also harmonic, we can  transform this terms for \emph{on-shell spin 4 gauge field} to the:
	\begin{eqnarray}
	&&-\frac{2}{3}  \varepsilon^{ \beta \gamma }_\mu  \partial_\beta h^{\nu \lambda \rho  \mu } \Box  J^{(4)}_{\nu \lambda \rho \gamma}
	- \partial^\beta \varepsilon^{ \gamma }_{\mu \nu}  h^{ \lambda \rho  \mu \nu} \Box  J^{(4)}_{\lambda \rho \beta \gamma} =
	\nonumber\\
	&&\frac{1}{12} \big\{  -8 \varepsilon^{ \beta \gamma }_\mu  \partial_\beta h^{\nu \lambda \rho  \mu }
	- 24\partial^\beta \varepsilon^{ \gamma }_{\mu \nu}  h^{ \lambda \rho  \mu \nu}
	-12\partial^\mu \varepsilon^{ \beta \gamma }_{\nu}  h^{ \lambda \rho  \mu \nu} \big\} \Box  J^{(4)}_{\nu \lambda \rho \gamma}
	\nonumber\\
	&&+\frac{1}{4}\delta_0 \big\{ h^{\beta \gamma }_{\mu \nu} h^{\lambda \rho \mu \nu}\big\}\Box  J^{(4)}_{\lambda \rho \beta \gamma} ,\label{5.13}
	\end{eqnarray}
	where we used again (\ref{5.2}) to integrate term in last line. Investigating expression in brackets of the second line of (\ref{5.13})
	we see that it is exactly trace of our spin 6 gauge field transformation introduced in (\ref{5.6})
	\begin{eqnarray}
	\delta_1 h^{\mu \nu \lambda \rho \alpha}_{\alpha}\Box  J^{(4)}_{\mu \nu \lambda \rho}= \Big\{ 8 \varepsilon^{\alpha \beta \rho}\partial_\alpha h^{\mu \nu \lambda}_\beta +12 \partial^\beta \varepsilon^{\lambda \rho}_{\alpha} h^{\mu \nu \alpha}_{\beta}+24 \partial^\mu \varepsilon ^{\nu \beta}_\alpha h^{\lambda \rho \alpha}_\beta \Big\}\Box  J^{(4)}_{\mu \nu \lambda \rho } .\label{5.14}
	\end{eqnarray}
	So we completely get rid of all reminders and obtained additional cubic interaction for trace of spin 6 field:
	\begin{equation}\label{5.15}
	\frac{1}{12}h^{\lambda \rho \beta \gamma \alpha}_{\alpha}\Box  J^{(4)}_{\lambda \rho \beta \gamma} ,
	\end{equation}
	and one more interacting term for second order on gauge field spin four interaction:
	\begin{equation}\label{5.16}
	L^{3}_{2}=-\frac{1}{4}h^{\beta \gamma }_{\mu \nu} h^{\lambda \rho \mu \nu}\Box  J^{(4)}_{\lambda \rho \beta \gamma} .
	\end{equation}
	Summing all $L^{i}_{2}, i=1,2,3$ we arrive to the quartic interaction (\ref{5.7last})
	
	\section*{Appendix B: A,B,C of generalized curvature and Christoffel symbols for spin $2\leq s \leq 4$ }
	\renewcommand{\theequation}{B.\arabic{equation}}\setcounter{equation}{0}
	To finalize our consideration the natural task should be consideration of the commutator of linear on gauge field gauge transformation (\ref{5.8}) obtained during construction of our quartic interaction. To do that we should first consider some exact relations for generalized Christoffel symbols and Curvatures introduced first time in  \cite{deWit:1979sib}. The alphabet of $s=4$ generalized curvature and Christoffel symbols are considered in details in \cite{Manvelyan:2007ey}. Here we present exact formulas for $s=2,3,4$ and derive nice reduction relations connected with antisymmetrization of possible  pairs of indices. So introducing gauge fields for spin $s\leq 4$
	\begin{equation}\label{7.1}
	h^{(4)}(x)_{\mu\nu\lambda\rho}, h^{(3)}(x)_{\mu\nu\lambda}, h^{(2)}(x)_{\mu\nu} .
	\end{equation}
	we can define the hierarchy of the Generalized Christoffel symbols and Curvature for spin 4:
	\begin{align}
	\Gamma^{(1)}_{\alpha;\mu\nu\lambda\rho}(h^{(4)})&=\partial_{\alpha}h^{(4)}_{\mu\nu\lambda\rho}-\partial_{(\mu}h^{(4)}_{\nu\lambda\rho)\alpha} ,\label{7.2}\\
	\Gamma^{(2)}_{\alpha\beta;\mu\nu\lambda\rho}(h^{(4)})&= \partial_{\alpha}\partial_{\beta}h^{(4)}_{\mu\nu\lambda\rho}-\frac{1}{2}\partial_{<\alpha}\partial_{(\mu}
	h^{(4)}_{\nu\lambda\rho)\beta>}+\partial_{(\mu}\partial_{\nu}h^{(4)}_{\lambda\rho)\alpha\beta}  ,\label{7.3}\\
	\Gamma^{(3)}_{\alpha\beta\gamma;\mu\nu\lambda\rho}(h^{(4)})&= \partial_{\alpha}\partial_{\beta}\partial_{\gamma}
	h^{(4)}_{\mu\nu\lambda\rho}-\frac{1}{3}\partial_{<\alpha}\partial_{\beta}\partial_{(\mu}
	h^{(4)}_{\nu\lambda\rho)\gamma>}+\frac{1}{3}\partial_{<\alpha}\partial_{(\mu}\partial_{\nu}
	h^{(4)}_{\lambda\rho)\beta\gamma>}\nonumber\\
	&-\partial_{(\mu}\partial_{\nu}\partial_{\lambda}
	h^{(4)}_{\rho)\alpha\beta\gamma} , \label{7.4}\\
	R^{(4)}_{\eta\alpha\beta\gamma;\mu\nu\lambda\rho}(h^{(4)})&=
	\Gamma^{(4)}_{\eta\alpha\beta\gamma;\mu\nu\lambda\rho}(h^{(4)})= \partial_{\eta}\partial_{\alpha}\partial_{\beta}\partial_{\gamma}
	h^{(4)}_{\mu\nu\lambda\rho}-\frac{1}{4}
	\partial_{<\eta}\partial_{\alpha}\partial_{\beta}\partial_{(\mu}
	h^{(4)}_{\nu\lambda\rho)\gamma>}\nonumber\\&+\frac{1}{6}
	\partial_{<\eta}\partial_{\alpha}\partial_{(\mu}\partial_{\nu}
	h^{(4)}_{\lambda\rho)\beta\gamma>}
	-\frac{1}{4}\partial_{<\eta}\partial_{(\mu}\partial_{\nu}\partial_{\lambda}
	h^{(4)}_{\rho)\alpha\beta\gamma>}+\partial_{\mu}\partial_{\nu}\partial_{\lambda}\partial_{\rho}
	h^{(4)}_{\eta\alpha\beta\gamma} ,\label{7.5}
	\end{align}
	with the following rule for zero order gauge transformation equipped by third rank symmetric tensor parameter:
	\begin{eqnarray}
	\delta^{(\varepsilon)}_{0}\Gamma^{(1)}_{\alpha;\mu\nu\lambda\rho}(h^{(4)})&=& -2\partial_{(\mu}\partial_{\nu}\varepsilon^{(3)}_{\lambda\rho)\alpha} ,\label{7.6}\\
	\delta^{(\varepsilon)}_{0}\Gamma^{(2)}_{\alpha\beta;\mu\nu\lambda\rho}(h^{(4)})&=& 3\partial_{(\mu}\partial_{\nu}\partial_{\lambda}\varepsilon^{(3)}_{\rho)\alpha\beta} ,\label{7.7}\\
	\delta^{(\varepsilon)}_{0}\Gamma^{(3)}_{\alpha\beta\gamma;\mu\nu\lambda\rho}(h^{(4)}) &=& -4\partial_{\mu}\partial_{\nu}\partial_{\lambda}\partial_{\rho}\varepsilon^{(3)}_{\alpha\beta\gamma} ,\label{7.8}\\
	\delta^{(\varepsilon)}_{0}R^{(4)}_{\eta\alpha\beta\gamma;\mu\nu\lambda\rho}(h^{(4)})&=&0 .\label{7.9}
	\end{eqnarray}
	The generalized curvature is invariant as it should be.
	
	Corresponding hierarchy of the Connections and Curvature can be defined  spin three:
	\begin{align}
	\Gamma^{(1)}_{\alpha;\mu\nu\lambda}(h^{(3)})&=\partial_{\alpha}h^{(3)}_{\mu\nu\lambda}-\partial_{(\mu}h^{(3)}_{\nu\lambda)\alpha} ,\label{7.10}\\
	\Gamma^{(2)}_{\alpha\beta;\mu\nu\lambda}(h^{(3)})&= \partial_{\alpha}\partial_{\beta}h^{(3)}_{\mu\nu\lambda}-\frac{1}{2}\partial_{<\alpha}\partial_{(\mu}
	h^{(3)}_{\nu\lambda)\beta>}+\partial_{(\mu}\partial_{\nu}h^{(3)}_{\lambda)\alpha\beta}  ,\label{7.11}\\
	R^{(3)}_{\alpha\beta\gamma;\mu\nu\lambda}(h^{(3)})&=\Gamma^{(3)}_{\alpha\beta\gamma;\mu\nu\lambda}(h^{(3)})= \partial_{\alpha}\partial_{\beta}\partial_{\gamma}
	h^{(3)}_{\mu\nu\lambda}-\frac{1}{3}\partial_{<\alpha}\partial_{\beta}\partial_{(\mu}
	h^{(3)}_{\nu\lambda)\gamma>}\nonumber\\&+\frac{1}{3}\partial_{<\alpha}\partial_{(\mu}\partial_{\nu}
	h^{(3)}_{\lambda)\beta\gamma>}-\partial_{\mu}\partial_{\nu}\partial_{\lambda}
	h^{(3)}_{\alpha\beta\gamma} . \label{7.12}
	\end{align}
	With corresponding gauge transformation with traceless symmetric second rank tensor parameter:
	\begin{eqnarray}
	\delta^{(\varepsilon)}_{0}\Gamma^{(1)}_{\alpha;\mu\nu\lambda}(h^{(3)})&=& -2\partial_{(\mu}\partial_{\nu}\varepsilon^{(2)}_{\lambda)\alpha} ,\label{7.13}\\
	\delta^{(\varepsilon)}_{0}\Gamma^{(2)}_{\alpha\beta;\mu\nu\lambda}(h^{(3)})&=& 3\partial_{\mu}\partial_{\nu}\partial_{\lambda}\varepsilon^{(2)}_{\alpha\beta} ,\label{7.14}\\
	\delta^{(\varepsilon)}_{0}R^{(3)}_{\alpha\beta\gamma;\mu\nu\lambda}(h^{(3)})&=&0 .\label{7.15}
	\end{eqnarray}
	And the same type relations for spin two:
	\begin{align}
	\Gamma^{(1)}_{\alpha;\mu\nu}(h^{(2)})&=\partial_{\alpha}h^{(2)}_{\mu\nu}-\partial_{(\mu}h^{(2)}_{\nu)\alpha} ,\label{7.16}\\
	R^{(2)}_{\alpha\beta;\mu\nu}(h^{(2)})&=\Gamma^{(2)}_{\alpha\beta;\mu\nu}(h^{(2)})= \partial_{\alpha}\partial_{\beta}h^{(2)}_{\mu\nu}-\frac{1}{2}\partial_{<\alpha}\partial_{(\mu}
	h^{(3)}_{\nu)\beta>}+\partial_{\mu}\partial_{\nu}h^{(2)}_{\alpha\beta}  ,\label{7.17}\\
	\delta^{(\varepsilon)}_{0}\Gamma^{(1)}_{\alpha;\mu\nu}(h^{(2)})&= -2\partial_{\mu}\partial_{\nu}\varepsilon^{(1)}_{\alpha} ,\label{7.18}\\
	\delta^{(\varepsilon)}_{0}R^{(2)}_{\alpha\beta;\mu\nu}(h^{(2)})&= 0 .\label{7.19}
	\end{align}
	General formulation for the general spin can be found in original paper \cite{deWit:1979sib} (see also \cite{Manvelyan:2007ey},\cite{Manvelyan:2007hv} for generalization to the $AdS$ space). Here in this Appendix we presented detailed formulas for spin $2\leq s \leq 4$ to prove reduction relation between different objects to use that for calculation f the commutator of two $\delta_{1}$ variations.
	
	To derive these relations we should choose one representative from each set of symmetrized indices  in   antisymmetrize one pair of indices from two sets of symmetrized indices in (\ref{7.4}) and compare with (\ref{7.11}) then we arrive to:
	\begin{align}
	\Gamma^{(3)}_{\alpha\beta[\gamma;\rho]\mu\nu\lambda}(h^{(4)})&=\frac{4}{3}
	\Gamma^{(2)}_{\alpha\beta;\mu\nu\lambda}(H^{(3)}_{[\gamma\rho]}) ,\label{7.20}\\
	\Gamma^{(2)}_{\alpha[\beta;\rho]\mu\nu\lambda}(h^{(4)})&=\frac{3}{2}
	\Gamma^{(1)}_{\alpha;\mu\nu\lambda}(H^{(3)}_{[\beta\rho]}) ,\label{7.21}\\
	\Gamma^{(1)}_{[\alpha;\rho]\mu\nu\lambda}(h^{(4)})&=2
	H^{(3)}_{[\alpha\rho];\mu\nu\lambda} ,\label{7.22}
	\end{align}
	where
	\begin{align}
	H^{(3)}_{[\gamma\rho];\mu\nu\lambda}&=\partial_{[\gamma}h^{(4)}_{\rho]m\nu\lambda}\label{7.23}
	\end{align}
	is first skew derivative of our spin four gauge field. So we see that antisymmetrized derivative pair behaves inert in symmetric construction of the Generalized Christoffel symbols and leads to the reduction relation (\ref{7.20})-(\ref{7.22}).
	
	The same type of antisymmetrization reduction relations exist for spin two-three case:
	\begin{align}
	\Gamma^{(2)}_{\alpha[\beta;\lambda]\mu\nu}(h^{(3)})&=\frac{3}{2}
	\Gamma^{(1)}_{\alpha;\mu\nu\lambda}(H^{(2)}_{[\beta\rho]}) ,\label{7.24}\\
	\Gamma^{(1)}_{[\alpha;\lambda]\mu\nu}(h^{(3)})&=2
	H^{(2)}_{[\alpha\lambda];\mu\nu} ,\label{7.25}\\
	H^{(2)}_{[\alpha\lambda];\mu\nu}&=\partial_{[\alpha}h^{(3)}_{\lambda]\mu\nu} .\label{7.26}
	\end{align}
	Then combining (\ref{7.20}) and (\ref{7.25}) we can derive second reduction formula for spin four third Christoffel symbol:
	\begin{align}
	\Gamma^{(3)}_{\alpha\big[\beta[\gamma;\rho]\lambda\big]\mu\nu}(h^{(4)})&=2
	\Gamma^{(1)}_{\alpha;\mu\nu}(H^{(2)}_{[\gamma\rho][\beta\lambda]}) ,\label{7.27}\\
	H^{(2)}_{[\gamma\rho][\beta\lambda];\mu\nu}&=\partial_{\big[\beta}\partial_{[\gamma}h^{(4)}_{\rho]\lambda\big] \mu\nu} .\label{7.28}
	\end{align}
	Gauge transformations for reduced Christoffels look like:
	\begin{eqnarray}
	&&\delta^{\varepsilon}_{0}\Gamma^{(2)}_{\beta\gamma;\nu\lambda\rho}(H^{(3)}_{[\eta\alpha]})=
	3\partial_{\nu}\partial_{\lambda}\partial_{\rho}
	E^{(2)}_{[\eta\alpha];\beta\gamma}(\varepsilon^{(3)}) ,\label{7.29}\\
	&&E^{(2)}_{[\eta\alpha];\beta\gamma}(\varepsilon^{(3)})= \partial_{[\eta}\varepsilon^{(3)}_{\alpha]\beta\gamma} ,\label{7.30}\\
	&&\delta^{\varepsilon}_{0}\Gamma^{(1)}_{\gamma;\lambda\rho}(H^{(2)}_{[\eta\alpha][\sigma\beta]})
	=-2\partial_{\lambda}\partial_{\rho}
	E^{(1)}_{[\eta\alpha][\sigma\beta]\gamma}(\varepsilon^{(3)}) ,\label{7.31}\\
	&&  E^{(1)}_{[\eta\alpha][\sigma\beta];\gamma}(\varepsilon)= \partial_{\big[\sigma}\partial_{[\eta}\varepsilon^{(3)}_{\alpha]\beta\big]\gamma} ,\label{7.32}
	\end{eqnarray}
	with corresponding gauge transformation for antisymmetrized derivatives of gauge field:
	\begin{eqnarray}
	&& \delta^{\varepsilon}_{0}H^{(3)}_{[\eta\alpha];\mu\nu\lambda}=\partial_{(\mu}E^{(2)}_{[\eta\alpha];\nu\lambda)} ,\label{7.33} \\
	&& \delta^{\varepsilon}_{0}H^{(2)}_{[\eta\alpha][\sigma\beta];\nu\lambda}=\partial_{(\nu}E^{(1)}_{[\eta\alpha][\sigma\beta];\lambda)} . \label{7.34}
	\end{eqnarray}
	Reduction relations work also for gauge invariant  curvatures
	\begin{align}
	R^{(4)}_{\eta\alpha\beta[\gamma;\rho]\mu\nu\lambda}(h^{(4)})&=\frac{5}{4}
	R^{(3)}_{\eta\alpha\beta;\mu\nu\lambda}(H^{(3)}_{[\gamma\rho]}) ,\label{7.35}\\
	R^{(3)}_{\eta\alpha[\beta;\lambda]\mu\nu}(h^{(3)})&=\frac{4}{3}R^{(2)}_{\eta\alpha;\mu\nu}(H^{(2)}_{[\beta\lambda]}) .\label{7.36}
	\end{align}
	Then we end this Appendix with some Bianchi Identity like formulas which we widely use in our calculation in this paper:
	\begin{align}
	\partial_{\eta}\Gamma^{(3)}_{\alpha\beta\gamma;\mu\nu\lambda\rho}(h^{(4)})&= \frac{1}{4}\partial_{(\mu}\Gamma^{(3)}_{|\alpha\beta\gamma|;\nu\lambda\rho)\eta}(h^{(4)})
	+R^{(4)}_{\eta\alpha\beta\gamma;\mu\nu\lambda\rho}(h^{(4)}) ,\label{7.37}\\
	R^{(4)}_{(\eta\alpha\beta\gamma;\mu)\nu\lambda\rho}(h^{(4)})  & =0 , \quad R^{(4)}_{\eta\alpha\beta\gamma;\mu\nu\lambda\rho}(h^{(4)})=R^{(4)}_{\mu\nu\lambda\rho;\eta\alpha\beta\gamma}(h^{(4)})  ,\label{7.38}\\
	\partial_{[\delta}R^{(4)}_{\eta]\alpha\beta\gamma;\mu\nu\lambda\rho}&= \frac{1}{5}\partial_{(\mu}R^{(4)}_{\nu\lambda\rho)[\delta;\eta]\alpha\beta\gamma} .\label{7.39}
	\end{align}
	
	\section*{Appendix C: Derivation of (\ref{6.13})-(\ref{6.15})}
	\renewcommand{\theequation}{C.\arabic{equation}}\setcounter{equation}{0}
	So now we can analyze remaining two terms in (\ref{6.10}).
	Taking $\delta_{0}$ variation of the (\ref{6.3}) we see easily that
	\begin{align}\label{C.13}
	&\Gamma^{(3)}_{\alpha\beta\gamma;\mu\nu\lambda\rho}(h)
	\delta^{(\omega)}_{0}\Lambda^{\alpha\beta\gamma}(\varepsilon,h)=
	\left[\varepsilon^{\delta\sigma\eta}\partial_{\delta}\partial_{\sigma}\partial_{\eta}
	\omega^{\alpha\beta\gamma}+\varepsilon^{\delta\sigma\eta}
	\partial^{\alpha}\partial^{\beta}\partial^{\gamma}\omega_{\delta\sigma\eta} +\frac{3}{2}\partial^{\alpha}\varepsilon^{\delta\sigma\eta}\partial_{\delta}\delta_{0}^{(\omega)}h_{\sigma\eta}^{\,\,\,\beta\gamma}
	\right.\nonumber\\
	&\left.-\partial^{\alpha}\varepsilon^{\delta\sigma\eta}
	\partial^{\beta}\delta_{0}^{(\omega)}h^{\gamma}_{\delta\sigma\eta}+\partial^{\alpha}\partial^{\beta}
	\varepsilon^{\delta\sigma\eta}\delta_{0}^{(\omega)}
	h_{\,\,\delta\sigma\eta}^{\gamma}\right]
	\Gamma^{(3)}_{\alpha\beta\gamma;\mu\nu\lambda\rho}(h)
	-(\varepsilon\leftrightarrow\omega) .
	\end{align}
	The first term in r.h.s of (\ref{6.10}) can be split into the sum of the following terms:
	\begin{align}\label{C.14}
	&\varepsilon^{\alpha\beta\gamma}\Gamma^{(3)}_{\alpha\beta\gamma;\mu\nu\lambda\rho}
	(\delta^{(\omega)}_{1}h)\sim \varepsilon^{\delta\sigma\eta}\partial_{\delta}\partial_{\sigma}\partial_{\eta}
	\omega^{\alpha\beta\gamma} \Gamma^{(3)}_{\alpha\beta\gamma;\mu\nu\lambda\rho}(h)
	+I^{(2)}_{\mu\nu\lambda\rho}(\varepsilon,\omega,h)+ I^{(1)}_{\mu\nu\lambda\rho}(\varepsilon,\omega,h)\nonumber\\
	&4\partial_{\mu}\partial_{\nu}\partial_{\lambda}\partial_{\rho}
	\varepsilon^{\alpha\beta\gamma}\omega^{\delta\sigma\eta}
	(\partial_{\delta}\partial_{\sigma}h_{\eta\alpha\beta\gamma}+\partial_{\alpha}
	\partial_{\beta}h_{\gamma\delta\sigma\eta}-\frac{3}{2}
	\partial_{\eta}\partial_{\alpha}h_{\delta\sigma\beta\gamma})-(\varepsilon\leftrightarrow\omega) ,
	\end{align}
	where
	\begin{eqnarray}
	I^{(2)}_{\mu\nu\lambda\rho}(\varepsilon,\omega,h)&&\sim 3\varepsilon^{\delta\sigma\eta}\partial_{\delta}\partial_{\sigma}
	\omega^{\alpha\beta\gamma}\partial_{\eta}
	\Gamma^{(3)}_{\alpha\beta\gamma;\mu\nu\lambda\rho}(h)+4\partial_{\mu}\varepsilon^{\delta\sigma\eta}\partial_{\delta}\partial_{\sigma}
	\omega^{\alpha\beta\gamma}
	\Gamma^{(3)}_{\alpha\beta\gamma;\eta\nu\lambda\rho}(h) ,\quad\qquad\label{C.15} \\
	I^{(1)}_{\mu\nu\lambda\rho}(\varepsilon,\omega,h)&&\sim 3\varepsilon^{\delta\sigma\eta}\partial_{\delta}
	\omega^{\alpha\beta\gamma}\partial_{\sigma}\partial_{\eta}
	\Gamma^{(3)}_{\alpha\beta\gamma;\mu\nu\lambda\rho}(h)+8\partial_{\mu}
	\varepsilon^{\delta\sigma\eta}\partial_{\delta}
	\omega^{\alpha\beta\gamma}
	\partial_{\sigma}\Gamma^{(3)}_{\alpha\beta\gamma;\eta\nu\lambda\rho}(h)\nonumber\\
	&&+6\partial_{\mu}\partial_{\nu}
	\varepsilon^{\delta\sigma\eta}\partial_{\delta}
	\omega^{\alpha\beta\gamma}\Gamma^{(3)}_{\alpha\beta\gamma;\sigma\eta\lambda\rho}(h) .
	\label{C.16}
	\end{eqnarray}
	Then using transformation rule (\ref{7.8}) to integrate $\Gamma^{(3)\alpha\beta\gamma}_{\qquad;\mu\nu\lambda\rho}(h)$ in r.h.s of (\ref{C.14}) we observe cancelation of the second line of (\ref{C.14}) with first line of (\ref{C.13}). So we arrive to the following preliminary result:
	\begin{eqnarray}
	[\delta^{(\omega)}_{1},\delta^{(\varepsilon)}_{1}]h_{\mu\nu\lambda\rho}&&\sim  \left[\varepsilon^{\delta\sigma\eta}\partial_{\delta}\partial_{\sigma}\partial_{\eta}
	\omega^{\alpha\beta\gamma}+T_{1}^{\alpha\beta\gamma}(\partial,\varepsilon,\omega)
	\right] \Gamma^{(3)}_{\alpha\beta\gamma;\mu\nu\lambda\rho}(h)\nonumber\\
	&&+I^{(2)}_{\mu\nu\lambda\rho}(\varepsilon,\omega,h)+ I^{(1)}_{\mu\nu\lambda\rho}(\varepsilon,\omega,h) -(\varepsilon\leftrightarrow\omega) ,\label{C.17}
	\end{eqnarray}
	where
	\begin{eqnarray}
	&& T_{1}^{\alpha\beta\gamma}=\frac{1}{2}\partial^{(\alpha}\varepsilon^{\delta\sigma\eta}
	\partial_{\delta}\delta_{0}^{(\omega)}h_{\sigma\eta}^{\,\,\,\beta\gamma)}
	-\frac{1}{6}\partial^{(\alpha}\varepsilon^{\delta\sigma\eta}
	\partial^{\beta}\delta_{0}^{(\omega)}h^{\gamma)}_{\delta\sigma\eta}+\frac{1}{3}
	\partial^{(\alpha}\partial^{\beta}
	\varepsilon^{\delta\sigma\eta}\delta_{0}^{(\omega)}
	h_{\,\,\delta\sigma\eta}^{\gamma)} \nonumber\\
	&&=\partial^{(\alpha}\varepsilon^{\delta\sigma\eta}\partial_{\delta}\partial_{\sigma}
	\omega^{\beta\gamma)}_{\eta}+\partial^{(\alpha}\partial^{\beta}
	\varepsilon^{\delta\sigma\eta}\partial_{\delta}
	\omega^{\gamma)}_{\sigma\eta}+\frac{1}{3}\partial^{(\alpha}\partial^{\beta}
	\varepsilon^{\delta\sigma\eta}\partial^{\gamma)}
	\omega_{\delta\sigma\eta}-\frac{1}{3}\partial^{(\alpha}\varepsilon^{\delta\sigma\eta}
	\partial^{\beta}\partial^{\gamma)}
	\omega_{\delta\sigma\eta} \nonumber\\\label{C.18}
	\end{eqnarray}
	is some additional input to  the symmetric rang three generalized tensorial bracket of two parameters $\varepsilon$ and $\omega$ with three derivatives:
	\begin{equation}\label{C.19}
	[\omega,\varepsilon]_{1}^{\alpha\beta\gamma} = \varepsilon^{\delta\sigma\eta}\partial_{\delta}\partial_{\sigma}\partial_{\eta}
	\omega^{\alpha\beta\gamma}+T_{1}^{\alpha\beta\gamma}(\partial,\varepsilon,\omega) .
	\end{equation}
	But it is not the full story about commutator because we still have unclassified transformations in two objects (\ref{C.15}) and (\ref{C.16}) of the second line of (\ref{C.17}).
	Our goal is to extract all transformation terms including more than three rang symmetric tensors as a combined parameter, correcting in parallel the third rank bracket (\ref{C.19}).
	Indeed using integration relations:
	\begin{align}
	&\partial_{\delta}\omega_{\alpha\beta\gamma}=\delta^{(\omega)}_{0} h_{\delta\alpha\beta\gamma}-\partial_{(\alpha}\omega_{\beta\gamma)\delta} ,\\
	&\partial_{\delta}\omega_{\alpha\beta\gamma}\dots\Gamma^{(3)\alpha\beta\gamma}_{\qquad;\mu\nu\lambda\rho}(h)
	=\big(\frac{3}{4}\partial_{[\delta}\omega_{\alpha]\beta\gamma}+\frac{1}{4}\delta^{(\omega)}_{0} h_{\delta\alpha\beta\gamma}\big)\dots\Gamma^{(3)\alpha\beta\gamma}_{\qquad;\mu\nu\lambda\rho}(h) ,
	\end{align}
	and after long calculation we can transform the sum of  $I^{(1)}$ and $I^{(2)}$ in the following way:
	\begin{align}
	&I^{(1)}_{\mu\nu\lambda\rho}(\varepsilon,\omega,h)
	+I^{(2)}_{\mu\nu\lambda\rho}(\varepsilon,\omega,h) \sim \tilde{I}^{(1)}_{\mu\nu\lambda\rho}(\varepsilon,\omega,H^{(3)})
	+\tilde{I}^{(2)}_{\mu\nu\lambda\rho}(\varepsilon,\omega,H^{(3)})\nonumber\\ &-\frac{1}{4}\partial^{\alpha}\partial^{\beta}\varepsilon^{\delta\sigma\eta}\delta_{0}^{(\omega)}
	h_{\,\,\delta\sigma\eta}^{\gamma}\Gamma^{(3)}_{\alpha\beta\gamma;\mu\nu\lambda\rho}(h)+3\varepsilon^{\delta\sigma\eta}\partial_{\delta}
	\partial_{\sigma}\omega^{\alpha\beta\gamma}
	R^{(4)}_{\eta\alpha\beta\gamma;\mu\nu\lambda\rho}(h) \nonumber\\
	&+\frac{9}{20}\varepsilon_{\delta}^{\sigma\eta}
	\partial^{[\delta}\omega^{\alpha]\beta\gamma}\Big[
	\partial_{(\sigma}R^{(4)}_{\eta\alpha\beta\gamma);\mu\nu\lambda\rho}(h)+
	\partial_{[\sigma}R^{(4)}_{\alpha]\beta\gamma\eta);\mu\nu\lambda\rho}(h)+ 2\partial_{[\sigma}R^{(4)}_{\beta]\gamma\alpha\eta);\mu\nu\lambda\rho}(h)
	\Big] ,\label{C.20}
	\end{align}
	where $ R^{(4)}_{\eta\alpha\beta\gamma;\mu\nu\lambda\rho}(h)$ is gauge invariant  spin  four linearized curvature (\ref{7.5}) with two set of four symmetrized indices\footnote{Here in last line of (\ref{C.20}) we separated symmetric and antisymmetric part of derivative of Generalized Curvature } and:
	\begin{align}
	\tilde{I}^{(2)}_{\mu\nu\lambda\rho}(\varepsilon,\omega,H^{(3)})&=\frac{3}{4}\Big[\frac{3}{4}\varepsilon^{\eta\sigma}_{\delta}\partial_{\sigma}
	\partial^{[\delta}\omega^{\alpha]\beta\gamma}\partial_{(\mu}
	\Gamma^{(2)}_{\beta\gamma;\nu\lambda\rho)}(H^{(3)}_{[\eta\alpha]})+\partial_{(\mu}\varepsilon^{\eta\sigma}_{\delta}\partial_{\sigma}
	\partial^{[\delta}\omega^{\alpha]\beta\gamma}
	\Gamma^{(2)}_{\beta\gamma;\nu\lambda\rho)}(H^{(3)}_{[\eta\alpha]})\Big] ,
	\quad\qquad\label{C.22} \\
	\tilde{I}^{(1)}_{\mu\nu\lambda\rho}(\varepsilon,\omega,H^{(3)})&=\frac{3}{4}\Big[\frac{3}{4}\varepsilon^{\eta\sigma}_{\delta}
	\partial^{[\delta}\omega^{\alpha]\beta\gamma}\partial_{\sigma}\partial_{(\mu}
	\Gamma^{(2)}_{\beta\gamma;\nu\lambda\rho)}(H^{(3)}_{[\eta\alpha]})+
	2\partial_{(\mu}\varepsilon^{\eta\sigma}_{\delta}
	\partial^{[\delta}\omega^{\alpha]\beta\gamma}
	\partial_{\sigma}\Gamma^{(2)}_{\beta\gamma;\nu\lambda\rho)}(H^{(3)}_{[\eta\alpha]})\nonumber\\
	&+\frac{3}{2}\partial_{(\mu}\partial_{\nu}
	\varepsilon^{\eta\sigma}_{\delta}
	\partial^{[\delta}\omega^{\alpha]\beta\gamma}
	\Gamma^{(2)}_{\beta\gamma;\sigma\lambda\rho)}(H^{(3)}_{[\eta\alpha]})\Big] ,
	\label{C.23}
	\end{align}
	where $\Gamma^{(2)}_{\beta\gamma;\nu\lambda\rho}(H^{(3)}_{[\eta\alpha]})$  is defined in (\ref{7.20}) as second Christoffel Symbol for third rank symmetric tensor $H^{(3)}_{[\eta\alpha];\beta\gamma\rho}$ with additional antisymmetric pair of indices. This tensor is actually curl of our spin four field (\ref{7.23}) and second Christoffel Symbol is constructed as usual for spin three field:
	\begin{eqnarray}
	&&  \Gamma^{(2)}_{\beta\gamma;\nu\lambda\rho}(H^{(3)}_{[\eta\alpha]})=
	\partial_{\beta}\partial_{\gamma}H^{(3)}_{[\eta\alpha];\beta\gamma\rho}
	-\frac{1}{2}\partial_{<\beta}\partial_{(\nu}H^{(3)}_{[\eta\alpha];\gamma\rho)\gamma>}
	+\partial_{(\nu}\partial_{\lambda}H^{(3)}_{[\eta\alpha];\rho)\beta\gamma} ,\quad\quad\label{C.25} \\
	&&\delta^{\varepsilon}_{0}\Gamma^{(2)}_{\beta\gamma;\nu\lambda\rho)}(H^{(3)}_{[\eta\alpha]})=
	3\partial_{\nu}\partial_{\lambda}\partial_{\rho}
	E^{(2)}_{[\eta\alpha];\beta\gamma}(\varepsilon) ,\label{C.26}\\
	&& E^{(2)}_{[\eta\alpha];\beta\gamma}(\varepsilon)= \partial_{[\eta}\varepsilon_{\alpha]\beta\gamma} .\label{C.27}
	\end{eqnarray}
	And we see that on this stage our commutator can be written in the following form:
	\begin{align}
	&[\delta^{(\omega)}_{1},\delta^{(\varepsilon)}_{1}]h_{\mu\nu\lambda\rho}\sim [\omega,\varepsilon]_{2}^{\alpha\beta\gamma}\Gamma^{(3)}_{\alpha\beta\gamma;\mu\nu\lambda\rho}
	(\delta^{(\omega)}_{1}h)+3\varepsilon^{\delta\sigma\eta}\partial_{\delta}
	\partial_{\sigma}\omega^{\alpha\beta\gamma}
	R^{(4)}_{\eta\alpha\beta\gamma;\mu\nu\lambda\rho}(h)\nonumber\\
	&+\frac{9}{20}\varepsilon_{\delta}^{\sigma\eta}
	\partial^{[\delta}\omega^{\alpha]\beta\gamma}\Big[
	\partial_{(\sigma}R^{(4)}_{\eta\alpha\beta\gamma);\mu\nu\lambda\rho}(h)
	+ \frac{1}{5}\partial_{(\mu}R^{(4)}_{\nu\lambda\rho)[\sigma;\alpha]\beta\gamma\eta}(h)+ \frac{2}{5}\partial_{(\mu}R^{(4)}_{\nu\lambda\rho)[\sigma;\beta]\gamma\alpha\eta}(h)\Big]\nonumber\\
	&+ \tilde{I}^{(1)}_{\mu\nu\lambda\rho}(\varepsilon,\omega,H^{(3)})+\tilde{I}^{(2)}_{\mu\nu\lambda\rho}(\varepsilon,\omega,H^{(3)})-(\varepsilon \leftrightarrow \omega) ,\label{C.28}
	\end{align}
	where in first line we define next deformation of spin four gauge symmetry parameter:
	\begin{eqnarray}
	&& [\omega,\varepsilon]_{2}^{\alpha\beta\gamma} = [\omega,\varepsilon]_{1}^{\alpha\beta\gamma}
	-\frac{1}{12}\partial^{(\alpha}\partial^{\beta}\varepsilon^{\delta\sigma\eta}\delta_{0}^{(\omega)}
	h_{\,\,\delta\sigma\eta}^{\gamma)} ,\label{C.29}
	\end{eqnarray}
	and in second line of (\ref{C.28}) we used Bianchi identity (\ref{7.39}).
	Then we can perform  one more step and extract dependence from symmetry parameter with two pair of antisymmetrized indices. To do that we use another cycle of gauge field integration from derivative of gauge parameter:
	\begin{align}
	&\partial_{\sigma}\partial_{[\delta}\omega_{\alpha]\beta\gamma}=\delta^{(\omega)}_{0} \partial_{[\delta}h_{\alpha]\beta\gamma\sigma}-\partial_{(\sigma}\partial_{[\delta}\omega_{\alpha]\beta\gamma)} ,\label{int1} \\
	&\partial_{\sigma}\partial_{[\delta}\omega_{\alpha]\beta\gamma}\dots\Gamma^{(2)\beta\gamma}_{\qquad;\nu\lambda\rho}(H^{(3)})
	=\big(\frac{2}{3}\partial_{\big[\sigma}\partial_{[\delta}\omega_{\alpha]\beta\big]\gamma}+\frac{1}{3}\delta^{(\omega)}_{0} \partial_{[\delta}h_{\alpha]\beta\gamma\sigma}\big)\dots\Gamma^{(2)\beta\gamma}_{\qquad;\nu\lambda\rho}(H^{(3)}) .\label{int2}
	\end{align}
	This leads to the following transformation:
	\begin{align}
	&\tilde{I}^{(1)}_{\mu\nu\lambda\rho}(\varepsilon,\omega,H^{(3)})
	+\tilde{I}^{(2)}_{\mu\nu\lambda\rho}(\varepsilon,\omega,H^{(3)}) \sim \tilde{\tilde{I}}^{(1)}_{\mu\nu\lambda\rho}(\varepsilon,\omega,H^{(2)})
	+\tilde{\tilde{I}}^{(2)}_{\mu\nu\lambda\rho}(\varepsilon,\omega,H^{(3)})\nonumber\\ &  -\frac{3}{16}\partial^{[\alpha}\varepsilon^{\delta]\sigma\eta}\delta^{(\omega)}_{0}
	\partial_{[\delta}h^{\gamma}_{\beta]\sigma\eta}
	\Gamma^{(3)\beta}_{\,\,\alpha\,\,\,\gamma;\mu\nu\lambda\rho}(h) \nonumber\\
	& -\frac{3}{2}\partial_{(\mu}\varepsilon_{\delta}^{\sigma\eta} \partial^{[\delta}\omega^{\alpha]\beta\gamma}R^{(3)}_{\nu\lambda\rho);\sigma\beta\gamma}(H^{3}_{[\eta\alpha]})-\frac{3}{16}\varepsilon_{\delta}^{\sigma\eta}
	\partial^{[\delta}\omega^{\alpha]\beta\gamma}\partial_{(\mu}R^{(3)}_{\nu\lambda\rho);\sigma\beta\gamma}(H^{3}_{[\eta\alpha]}) ,\label{bracket}
	\end{align}
	where
	\begin{align}
	\tilde{\tilde{I}}^{(2)}_{\mu\nu\lambda\rho}(\varepsilon,\omega,H^{(3)}) &= \frac{1}{2}\Big[\frac{3}{4}\varepsilon^{\eta}_{\delta\sigma}\partial^{\big[\sigma}
	\partial^{[\delta}\omega^{\alpha]\beta\big]\gamma}\partial_{(\mu}
	\Gamma^{(2)}_{\beta\gamma;\nu\lambda\rho)}(H^{(3)}_{[\eta\alpha]})+\partial_{(\mu}
	\varepsilon^{\eta}_{\delta\sigma}\partial^{\big[\sigma}
	\partial^{[\delta}\omega^{\alpha]\beta\big]\gamma}
	\Gamma^{(2)}_{\beta\gamma;\nu\lambda\rho)}(H^{(3)}_{[\eta\alpha]})\Big] ,
	\quad\qquad\label{C.30} \\
	\tilde{\tilde{I}}^{(1)}_{\mu\nu\lambda\rho}(\varepsilon,\omega,H^{(2)})&=\frac{3}{4}\Big[\frac{1}{2}\varepsilon^{\eta\sigma}_{\delta}
	\partial^{[\delta}\omega^{\alpha]\beta\gamma}\partial_{(\mu}\partial_{\nu}
	\Gamma^{(1)}_{\gamma;\lambda\rho)}(H^{(2)}_{[\eta\alpha][\sigma\beta]})+
	\frac{2}{3}\partial_{(\mu}\varepsilon^{\eta\sigma}_{\delta}
	\partial^{[\delta}\omega^{\alpha]\beta\gamma}
	\partial_{\nu}\Gamma^{(1)}_{\gamma;\lambda\rho)}(H^{(2)}_{[\eta\alpha][\sigma\beta]})\nonumber\\
	&+\partial_{(\mu}\partial_{\nu}
	\varepsilon^{\eta\sigma}_{\delta}
	\partial^{[\delta}\omega^{\alpha]\beta\gamma}
	\Gamma^{(1)}_{\gamma;\lambda\rho)}(H^{(2)}_{[\eta\alpha][\sigma\beta]})\Big] .
	\label{C.31}
	\end{align}
	Here in (\ref{C.31})  $\Gamma^{(1)}_{\gamma;\lambda\rho}(H^{(2)}_{[\eta\alpha][\sigma\beta]})$ is defined in (\ref{7.27}) as first Christoffel Symbol for second rank symmetric tensor $H^{(2)}_{[\eta\alpha][\sigma\beta];\gamma\rho}(h)$ with additional two antisymmetric pair of indices. This tensor is second curl of our spin four field:
	\begin{equation}\label{C.32}
	H^{(2)}_{[\eta\alpha][\sigma\beta];\gamma\rho}(h)=
	\partial_{\big[\sigma}\partial_{[\eta}h_{\alpha]\beta\big]\gamma\rho} ,
	\end{equation}
	and Christoffel Symbol is constructed as usual for spin two field ignoring both antisymmetric pairs:
	\begin{eqnarray}
	&& \Gamma^{(1)}_{\gamma;\lambda\rho}(H^{(2)}_{[\eta\alpha][\sigma\beta]})=
	\partial_{\gamma}H^{(2)}_{[\eta\alpha][\sigma\beta];\lambda\rho}
	-\partial_{(\lambda}H^{(2)}_{[\eta\alpha][\sigma\beta];\rho)\gamma} ,\label{C.33} \\
	&&\delta^{\varepsilon}_{0}\Gamma^{(1)}_{\gamma;\lambda\rho}(H^{(2)}_{[\eta\alpha][\sigma\beta]})
	=-2\partial_{\lambda}\partial_{\rho}
	E^{(1)}_{[\eta\alpha][\sigma\beta];\gamma}(\varepsilon) ,\label{C.34}\\
	&&   E^{(1)}_{[\eta\alpha][\sigma\beta];\gamma}(\varepsilon)= \partial_{\big[\sigma}\partial_{[\eta}\varepsilon_{\alpha]\beta\big]\gamma} .\label{C.35}
	\end{eqnarray}
	
Now we can finish our proof in three steps:
	\begin{enumerate}
		\item From  second term in r.h.s. of (\ref{bracket}) we can extract final deformation of the spin four gauge symmetry parameter
		\begin{eqnarray}
		&& [\omega,\varepsilon]_{3}^{\alpha\beta\gamma} = [\omega,\varepsilon]_{2}^{\alpha\beta\gamma}-\frac{3}{16}\partial^{[\alpha}\varepsilon^{\delta]\sigma\eta}\delta^{(\omega)}_{0}
		(\partial_{\delta}h^{\beta\gamma}_{\sigma\eta}-\partial^{\beta}h^{\gamma}_{\delta\sigma\eta})\nonumber\\&&+\textit{symmetrization in}\,\,\, (\alpha\beta\gamma)\,,\label{C.36}
		\end{eqnarray}
		and this expression after some algebra produces (\ref{6.14}) and therefore first line of (\ref{6.13}).
		\item Combining first two line of (\ref{C.28}), last line of (\ref{bracket}) and using relation (\ref{7.35}) we obtain all curvature dependent terms in (\ref{6.13}) and (\ref{6.15}) .
		\item All other terms with mixed symmetry parameters in (\ref{6.15}) are just sum of (\ref{C.30}) and (\ref{C.31}).
	\end{enumerate}
	So we prove our final result for commutator (\ref{6.13})-(\ref{6.15}).

\end{document}